\documentclass{IEEEtran}
\usepackage{amsmath,amssymb,cite}
\usepackage[pdftex]{graphicx}
\usepackage[inline]{asymptote}

\newcommand{\bbR}{{\mathbb R}}         
\newcommand{\bbC}{{\mathbb C}}         
\newcommand{\bbT}{{\mathbb T}}         
\newcommand{\bbZ}{{\mathbb Z}}         
\newcommand{\bbA}{{\mathbb A}}         
\newcommand{\bbI}{{\mathbb I}}         

\newcommand{\sE}{{\mathsf E}}          
\newcommand{\sT}{{\mathsf T}}          

\newcommand{\cN}{{\mathcal N}}         
\newcommand{\cS}{{\mathcal S}}
\newcommand{\cT}{{\mathcal T}}

\newcommand{\cF}{{\mathcal F}}

\newcommand{\fL}{{\mathfrak L}}
\newcommand{\fP}{{\mathfrak P}}

\newcommand{\br}{{\boldsymbol r}}
\newcommand{\bv}{{\boldsymbol v}}
\newcommand{\bu}{{\boldsymbol u}}

\newcommand{\bn}{{\boldsymbol n}}
\newcommand{\be}{{\boldsymbol e}}
\newcommand{\bx}{{\boldsymbol x}}
\newcommand{\by}{{\boldsymbol y}} 
\newcommand{\bq}{{\boldsymbol q}} 
\newcommand{\bs}{{\boldsymbol s}}
\newcommand{\bzero}{{\boldsymbol 0}}
\newcommand{\bz}{{\boldsymbol z}} 
\newcommand{\bg}{{\boldsymbol g}}
\newcommand{\bnabla}{{\boldsymbol \nabla}}
\newcommand{\beps}{{\boldsymbol \varepsilon}}
\newcommand{\bom}{{\boldsymbol \omega}}
\newcommand{\btau}{{\boldsymbol \tau}}
\newcommand{\bone}{{\boldsymbol 1}}
\newcommand{\bnu}{{\boldsymbol \nu}}

\newcommand{\norm}[1]{\lVert#1\rVert}
\newcommand{\abs}[1]{\lvert#1\rvert}
\newcommand{\linspan}{\operatorname{span}}
\newcommand{\diag}{\operatorname{diag}}
\newcommand{\tr}{\operatorname{Tr}}
\renewcommand{\Re}{\operatorname{Re}}
\renewcommand{\Im}{\operatorname{Im}}

\title{Estimation and Registration on Graphs}
\author{Stephen D.~Howard, Douglas Cochran, William Moran, and Frederick R. Cohen
\thanks{This work was supported in part by the DARPA Sensor Topology for Minimalist Planning (SToMP) program, the Australian Research Council, and the Defence Science and Technology Organisation of Australia.

S. D. Howard is with the Defence Science and Technology Organisation, PO Box 1500, Edinburgh 5111 SA, Australia, Stephen.Howard@dsto.defence.gov.au.

D. Cochran is with the School of Mathematical and Statistical Sciences, Arizona State University, Tempe AZ 85287-1804 USA, cochran@asu.edu.

W. Moran is with the Department of Electrical and Electronic Engineering, University of Melbourne, Parkville 3010 VIC, Australia, b.moran@ee.unimelb.edu.au.

F. R. Cohen is with the Department of Mathematics, University of Rochester, Rochester NY 14627, USA, fcoh@math.rochester.edu.
}
}

\begin{document}
\maketitle

\begin{abstract}
A statistical framework is introduced for a broad class of problems involving synchronization or registration of data across a sensor network in the presence of noise.  This framework enables an estimation-theoretic approach to the design and characterization of synchronization algorithms. The Fisher information is expressed in terms of the distribution of the measurement noise and standard mathematical descriptors of the network's graph structure for several important cases.  This leads to maximum likelihood and approximate maximum-likelihood registration algorithms and also to distributed iterative algorithms that, when they converge, attain statistically optimal solutions. The relationship between optimal estimation in this setting and Kirchhoff's laws is also elucidated.
\end{abstract}

\section{Introduction}
\label{sec:introduction}

Registration of data across a network is an ubiquitous problem in distributed sensing. Over more than three decades, much effort has been expended on development of algorithms to provide time synchronization across a distributed network; e.g.,\cite{Lamport1978,Ganeriwal2003,Su2005,Giridhar2006,Noh2008}. Synchronization of this kind is important for distributed parallel processing as well as data fusion across a sensor network. It is typically the case that the network is not complete; i.e., each node does not communicate with every other node. A large fraction of algorithms described in the literature produce algorithms to minimize an error or objective function based on least squares, often within power or other resource constraints. Leaving aside the latter issue, the problem in this setting is to assign a clock adjustment to every node based on knowledge of the clock differences, generally noisy, between some pairs of nodes in the network. Even if clock difference measurements are available for every pair of nodes in the network, the presence of noise still raises consistency considerations; e.g., the true offsets must sum to zero around any closed cycle.

The domain of practical network synchronization problems is by no means limited to clock offsets, nor is the natural measurement space restricted to the real line. Individual nodes may possess multiple data to be registered cross the network, and the noise affecting such vector data may be correlated across its components. Further, the natural measurement space is often not a linear space. In phase synchronization, for example, typical data could be measurements of the phase differences between local oscillators at the nodes.  In this case, the natural measurement space is the circle $\bbT=\bbR/2\pi \bbZ$ rather then the real line $\bbR$. If several local oscillators are involved, measurements might lie on the torus $\bbT^n$. Another important practical example where the measurement space is a nonlinear multi-dimensional manifold is registration of local coordinate systems, for which the natural setting is the special orthogonal group $SO(3)$. Even in the context of clocks, if both offset and clock speed are adjustable locally the offsets are elements of the affine group $\bbA$.  These examples illustrate that practical problems can entail data on Lie groups that are compact (e.g., $\bbT$ or $SO(3)$), non-compact ($\bbR^n$), abelian ($\bbR^n$, $\bbT$), or non-abelian ($SO(3)$ or $\bbA$).

It is common to represent networks by graphical models. In this setting, the network nodes that provide data to be registered or synchronized are represented by vertices labeled with their associated parameters, such as local clock time or local coordinate system. Each pair of vertices corresponding to a pair of nodes that are in direct communication are joined by an edge. Information is shared between vertices along such edges, each of which is labeled by a noisy measurement of the difference between the parameters at the end vertices of the edge. The notion of difference between two parameter values depends on the algebraic structure of the parameter space. In $\bbR^n$, for example, it is defined by subtraction of vectors. In other spaces, which are assumed to be Lie groups, difference is defined in terms of the group operation. The estimation problem on which this paper focuses arises precisely because the difference values that label the edges are corrupted by noise in a sense that will be made precise later in the paper. As a consequence of this corruption, the edge labels in any graph with cycles will generally be inconsistent; i.e., the edge labels around closed cycles will not sum to zero, even though the true difference values must do so. For a connected graph, the desired synchronization should provide a set of consistent edge labels that can be used to determine a unique offset value to register any pair of vertices in the graph. If one vertex label is known, this is equivalent to assigning labels to all other vertices consistently throughout the graph.

The first goal of this paper is to frame a class of network synchronization problems encompassed by the graphical model just described in terms of statistical estimation theory.  The second goal is to derive and characterize maximum-likelihood estimators or approximations thereof for a significant subclass class of these problems, together with local algorithms that realize these estimators. Specifically, these estimation problems entail estimation of the vertex labels (parameters) from the edge labels (noisy data). This paper focuses upon two cases in which the noise models are classical: Gaussian noise on $\bbR^{d}$ and von Mises distributed noise on $\bbT$. Between them, these cases encompass most of the elements encountered in the situation where the measurements and noise are on abelian Lie groups, which is the most general setting addressed in detail in this paper. The non-abelian situation, which presents additional mathematical challenges, will be addressed in a sequel. 

In each form of the estimation problem treated here, the Fisher information and maximum likelihood estimator are derived in a closed forms that depend both on the distribution of the noise and on the structure of the graph in intuitively appealing ways. In each of these cases, it is observed that the maximum likelihood estimator is non-local in the sense that the offsets for a node relative to its neighbors cannot be estimated at that node using only information obtained from its neighbors. Nevertheless, it is possible to find iterative (gradient descent) algorithms that are local and do converge to the ML estimator when they converge, which happens very frequently in empirical tests. 

Interesting problems associated with local maxima of the likelihood function that arise in the setting of compact Lie groups are discussed in connection with the circle case treated here. The value of the Fisher information and functions thereof (e.g., its determinant) in designing networks that support accurate registration is also discussed.  

Structurally, Section \ref{sec:graphs} begins with a synopsis of the essential concepts of graph theory and the associated notation that will be used throughout the paper. Some relevant introductory material on Lie groups is also summarized in this section, and the mathematical framework for discussion of the abelian Lie group is set forth.  Section \ref{sec:gaussian} treats the case of Gaussian noise on $\bbR^d$ and Section \ref{sec:localRd} subsequently describes and analyzes the performance of local algorithms for this case. Section \ref{sec:circle} covers the case of von Mises noise on $\bbT$, followed by description and performance analysis for local algorithms for this situation and a discussion of critical points of the likelihood function. The general abelian Lie group case is presented in Section \ref{sec:abelian_Lie}, where it is shown to follow essentially the same pattern as the special cases presented earlier. Finally, some preliminary remarks on the important non-abelian Lie group setting appears in Section \ref{sec:wrapup} together with some concluding remarks.

\section{Graph Theory Preliminaries}
\label{sec:graphs}

The aspects of elementary graph theory needed in the remainder of this paper are covered in many standard references, such as \cite{Bollobas1998}. In what follows, a graph $\Gamma$ will be assumed to have a finite vertex set $V(\Gamma)$. The collection of edges in $\Gamma$ will be denoted by $E(\Gamma)$, where the edge $e = (u,v)$ joins vertices $u$ and $v$ . If the graph is directed then edge $(u,v)$ starts at $u$ and ends at $v$. If the graph is not directed then $(u,v)\equiv (v,u)$. Unless otherwise noted, graphs in this paper will be directed, and the values labeling the edges will correspond to the difference between the label on the vertex at the end of the edge and the label on the vertex at the start of the edge. At several places in the paper, however, the graph orientation is irrelevant.  

In the following definitions of spaces of functions on $V(\Gamma)$ and $E(\Gamma)$, the functions are assumed to be real-valued, as in the clock synchronization example.  Later these definitions will be extended to cover other possible parameter spaces. The \emph{vertex space} $C_0(\Gamma)$ of a graph $\Gamma$ is the real vector space of functions $V(\Gamma)\rightarrow\bbR$; elements of $C_0(\Gamma)$ are vectors of real numbers indexed by the vertices. Similarly the \emph{edge space} $C_1(\Gamma)$ is the real vector space of functions $E(\Gamma)\rightarrow \bbR$.  The number of vertices in $\Gamma$ will be denoted by $|V(\Gamma)|=n$ and the number of edges by $|E(\Gamma)|=m$. It will be convenient to fix an ordering on the sets $V$ and $E$ in order to construct bases for $C_0(\Gamma)$ and $C_1(\Gamma)$. Denoting $V(\Gamma) = \{v_1,\ldots, v_n\}$, a basis for $C_0(\Gamma)$ is obtained by defining functions $\bv_1, \cdots, \bv_n$ according to
\begin{equation}
\label{eq:sbv}
\bv_i(v_j) = \delta_{ij}, \quad \text{$i,j = 1, \ldots, n$}
\end{equation}
so that any element of $C_0(\Gamma)$ can be written
\begin{equation*}
\bx = \sum_{i=1}^n x_i \bv_i
\end{equation*} 
$C_0(\Gamma)$ is an inner product space with inner product $\left< \cdot,\cdot\right>_{C_0}$ such that $\left< \bv_i,\bv_j\right>_{C_0} = \delta_{ij}$, so that $\bv_1,\cdots,\bv_n$ becomes an orthonormal basis of the real inner product space. The linear map $A: C_0(\Gamma)\rightarrow C_0(\Gamma)$ defined by
\begin{equation*}
A \bv_\ell = \sum_{v_j \sim v_\ell} \bv_j
\end{equation*}
where $v_j \sim v_\ell$ indicates that there is an edge connecting $v_j$ and $v_\ell$, is called the (undirected) \emph{adjacency} map. The corresponding matrix in the $\bv_\ell$ basis is called the adjacency matrix.

The degree $d_v$ of a vertex $v$ is the number of edges either starting or ending at $v$, and the linear map $N:C_0(\Gamma)\rightarrow C_0(\Gamma)$ defined by
\begin{equation*}
N \bv_\ell = d_{v_\ell}\bv_\ell, 
\end{equation*}
is called the degree map. The corresponding matrix in the $\bv_j$ basis is called the degree matrix and is diagonal with the degrees of the vertices of $\Gamma$ on its
main diagonal. The Kirchhoff map (unnormalized Laplacian) of $\Gamma$ is defined by
\begin{equation*}
L = N-A\ :\; C_0(\Gamma)\to C_1(\Gamma).
\end{equation*}
$L$ is positive semidefinite and the dimension of its zero eigenspace (null space) is the number of connected components of $\Gamma$.

Indexing the edges of $\Gamma$ as $E(\Gamma) = \{e_1,\cdots, e_m\}$, a basis for $C_1(\Gamma)$ can be constructed in similar fashion to the one for $C_{0}(\Gamma)$. Specifically, define functions $\be_1,\ldots,\be_n$ by
\begin{equation}
\label{eq:sbe}
\be_i(e_j) = \delta_{ij}, \quad \text{$i,j = 1, \ldots, m$}
\end{equation}
Then any element of $C_1(\Gamma)$ can be written
\begin{equation*}
\bx = \sum_{i=1}^n x_i \be_i
\end{equation*} 
Defining an inner product $\langle \cdot,\cdot\rangle_{C_1}$ such that $\langle
\be_i,\be_j\rangle_{C_1} = \delta_{ij}$ makes $C_1(\Gamma)$ into an inner product space.
Linear \emph{source} and \emph{target} maps, respectively $s,t :C_1(\Gamma)\rightarrow C_0(\Gamma)$ are defined by
\begin{align*}
s(\be_j) = \bu \quad \text{and} \quad t(\be_j) = \bv
\end{align*}
where $e_j=(u,v)$.  Finally the (directed) \emph{incidence} (or boundary) map
of $\Gamma$ is $D :C_1(\Gamma)\rightarrow C_0(\Gamma)$, where $D = t -
s$. Certain concepts will be needed from homology theory, and will be
introduced informally as required. The map 
$D$ is, in homological terms, a {\em boundary operator}, which applied to any edge $e=(u,v)$ gives
\begin{equation*}
D(\be) = t(\be)-s(\be) = \bv-\bu.
\end{equation*}
The Kirchhoff map can be written in terms of the incidence map as $L=DD^{\sT}$, where the adjoint map $D^{\sT}: C_0(\Gamma)\rightarrow C_1(\Gamma)$ is the \emph{coboundary} operator defined by
\begin{equation*}
D^{\sT}(\bv) = \sum_{e_j: t(e_j)=v} \be_j - \sum_{e_j: s(e_j)=v} \be_j.
\end{equation*}
\begin{figure}[h]
\begin{center}
\begin{asy}
size(5cm);
draw(Label("$D$", MidPoint, N), (0,0.1){NE}..{SE}(2,0.1), Arrow);
label("$C_1(\Gamma)$",(0,0),W);
draw(Label("$D^{\sT}$", MidPoint, S), (2,-0.1){SW}..{NW}(0,-0.1), Arrow);
label("$C_0(\Gamma)$",(2,0),E);
\end{asy}
\end{center}
\end{figure}

The \emph{cycle space} $Z(\Gamma)$ is defined as follows. A \emph{cycle} is a closed path in $\Gamma$; that is, a sequence of vertices $\fL=v_1v_2v_3\ldots v_q$ in $\Gamma$ where $v_{i}$ is adjacent to $v_{i+1}$ for $i=1,\ldots q$ and $v_{q}$ is adjacent to $v_{1}$.  The corresponding element of $Z(\Gamma)$ is a function $\bz_{\fL}\in C_1(\Gamma)$ given by
\begin{equation}\label{eq:cycle1}
\bz_{\fL}(e_j) = \left\{ \begin{array}{rl}
1 & \text{if $e_j\in \fL$ and $e_j$ is oriented as $\fL$}\\
-1 & \text{if $e_j\in \fL$ and $e_j$ is oriented opposite to $\fL$} \\
0 & \text{otherwise}
	\end{array}\right.
\end{equation}
$Z(\Gamma)$ is the linear subspace of $C_1(\Gamma)$ spanned by the $\bz_{\fL}$ as $\fL$ runs over all cycles in $\Gamma$. The cycle space is exactly the kernel of $D$; that is, for all $\bz\in Z(\Gamma)$,
\begin{equation}
\label{eq:1}
D\bz = 0,
\end{equation}
and every $\bz\in C_{1}(\Gamma)$ satisfying (\ref{eq:1}) is a linear
combination of $\bz_{\fL}$.  This condition implies that, for $\bz\in
Z(\Gamma)$, the oriented sum of the values on the set of edges meeting at any
of the vertices is zero; i.e.,
\begin{equation}
\label{eq:kirchhoff}
\sum_{e_j: t(e_j)=v} z_j = \sum_{e_j: s(e_j)=v} z_j,
\end{equation}
for all $v\in V(\Gamma)$. This, of course, is a statement of \emph{Kirchhoff's
  current law}. For a graph with $k$ connected components the dimension of
$Z(\Gamma)$ is the first Betti number of the graph; i.e., $\dim Z(\Gamma) =
m-n+k$.

A second subspace of $C_1(\Gamma)$ that arises in the development to follow is
the cocycle space (cut set) $U(\Gamma)$, an element of which is defined by fixing a partition $\fP$ of the vertex set $V(\Gamma)$ into two disjoint sets; i.e., $V(\Gamma) = V_1 \cup V_2$. With respect to this partition, define a vector $\bom_\fP \in U(\Gamma)$ to be
\begin{equation}
\label{eq:cuts}
\bom_\fP(e_j) = \left\{ \begin{array}{rl}1 &\text{if $e_j$ joins $V_1$ to $V_2$}\\
	-1 &\text{if $e_j$ joins $V_2$ to $V_1$}\\
	0 & \text{otherwise}
	\end{array}\right. \nonumber
\end{equation}
$U(\Gamma)$ is the linear subspace of $C_1(\Gamma)$ spanned by the $\bom_\fP$ as $\fP$ runs over all partitions of $V(\Gamma)$. 

The cocycle space is exactly the orthogonal complement of $Z(\Gamma)$ in $C_1(\Gamma)$. Thus, for any $\bz\in Z(\Gamma)$ and any $\bom\in U(\Gamma)$,
\begin{equation}\label{eq:cocycle1}
\langle\bz,\bom\rangle_{C_1} = 0
\end{equation}
This implies that every $\bom=\sum_{j}\omega_{j}\be_{j}\in U(\Gamma)$ satisfies
\begin{equation*}
\sum_{e_j\in \fL} (-1)^{\sigma_j} \omega_j = 0,
\end{equation*}
for all cycles $\fL$, where $\sigma_j = 0$ if $e_j$ is oriented as $\fL$ and $\sigma_j = 1$ if $e_j$ is oriented opposite to $\fL$. This is \emph{Kirchhoff's voltage law}. Furthermore, every vector $\bom\in C_1(\Gamma)$ can be uniquely decomposed as
\begin{equation*}
\bx = \bom + \bz, \quad \text{$\bom\in U(\Gamma)$ and $\bz \in Z(\Gamma)$}
\end{equation*}
In other words, $C_1(\Gamma)=Z(\Gamma)\oplus U(\Gamma)$. It is customary in the mathematical literature (e.g., \cite{Smale1972}) not to choose a basis to identify the $C_i(\Gamma)\ (i=0,1)$ with their duals and to regard boundary and coboundary maps as being on different spaces. It is convenient here to identify them.

The cocycle space $U(\Gamma)$ is also the image of $C_0(\Gamma)$ under the coboundary operator, $U(\Gamma) = \text{Im}(D^\sT)$. The kernel of $D^\sT$ is the space of locally constant functions on $V(\Gamma)$; i.e., functions that are constant on the vertices of connected components. In the development to follow, there is no loss of generality in working on different connected components separately.  So from this point $\Gamma$ will be assumed to be connected, and hence the kernel of $D^\sT$ is $\linspan\{\bone\}$ where
$\bone$ denotes the unit constant function on $V(\Gamma)$. With this connectedness assumption, every $\bx\in C_0(\Gamma)$ can be decomposed uniquely as
\begin{equation}
\label{eq:vertexdecomp}
\bx = D\bom + \alpha\bone,
\end{equation}
where $\bom\in U(\Gamma)$ and $\alpha\in\bbR$.  Formally, this decomposition is orthogonal since the kernel of $D^{\sT}$ is orthogonal to the image of $D$. 

A spanning tree $S$ in $\Gamma$ is a graph with $V(S)=V(\Gamma)$ such that every pair of vertices is joined by exactly one path in $S$. Equivalently, $S$ is a maximal subtree of $\Gamma$. Kirchhoff's matrix tree theorem implies that the number of spanning trees $t(\Gamma)$ in $\Gamma$ is equal to the absolute value of any cofactor of $L(\Gamma)$; i.e., $t(\Gamma)$ is the modulus of the product of the $n-1$ largest eigenvalues of $L(\Gamma)$. 

In graph theory, one often encounters maps $W:C_1(\Gamma)\rightarrow C_1(\Gamma)$ that describe some weighting on the edges of $\Gamma$. In the standard basis, such a map has diagonal matrix, and a weighted Laplacian can be defined by $\tilde L = D W D^\sT$. The matrix tree theorem can be generalized to state that the absolute value of any cofactor of $\tilde L$ is equal to 
\begin{equation*}
\sum_{S} \prod_{e \in S } W(e)
\end{equation*}
where the sum extends over all spanning trees of $\Gamma$.
%

\section{Gaussian Noise on $\bbR^d$}
\label{sec:gaussian}

This section provides precise formulations of estimation problems that arise in connection with registration on networks in situations where the parameter values at each node are real numbers or vectors in $\bbR^d$. The measurements of differences between parameter values at communicating nodes are corrupted by zero-mean additive Gaussian noise. The one-dimensional problem with independent noise on each measurement is treated first.  The value of explicit expressions for the Fisher information, its determinant, and maximum-likelihood estimators obtained for this case in network design problems is discussed.  This is followed by treatment of the multi-dimensional problem, in which both correlated and independent noise are considered.

\begin{figure}[h]
\begin{center}
\includegraphics[width=0.65\linewidth]{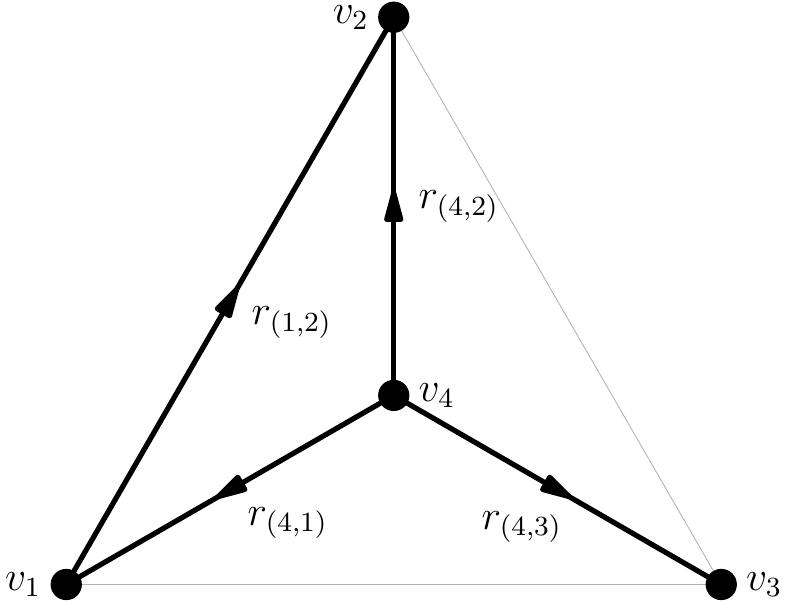}
\end{center}
\caption{Each edge $e$ of a directed graph $\Gamma$ is labeled with a value $r_e$ that
is the difference $\omega_e$ of labels at its boundary vertices plus a noise value $\varepsilon_e$.}
\end{figure}

\subsection{One-dimensional Gaussian problem}
\label{sec:GaussRind}

The situation in which the parameter space is the real line and the differences between communicating nodes are corrupted by zero-mean additive Gaussian noise is developed here in detail because it illustrates the approach that will be used in the more complicated cases that follow.
In this setting, the data vector $\br\in C_1(\Gamma)$, is a sum of a vector of true difference values and noise; i.e.,
\begin{equation}
\label{eq:GaussR}
\br = \bom + \beps
\end{equation}
where $\br$ and $\beps$ are in $C_1(\Gamma)$. Because the true difference values must sum to zero around any cycle, $\bom\in U(\Gamma)$. For the moment, assume that $\beps$ is jointly normal with covariance matrix $\sigma^2\bbI$; i.e., that random variables  $\varepsilon_i$ are independent and identically distributed (iid) with variance $\sigma^2$. With this assumption, the conditional probability density function for $\br$ given $\bom$ is then
\begin{eqnarray*}
p(\br | \bom) & = & \prod_{e_j\in E(\Gamma)} \frac{1}{\sqrt{2\pi\sigma^2}} \exp\left( -\frac{1}{2\sigma^2}(r_j - \omega_j)^2\right) \\
& = & (2\pi\sigma^2)^{-|E(\Gamma)|/2} \exp \left(-\frac{1}{2\sigma^2}\norm{\br - \bom}_{C_1}^2\right)
\end{eqnarray*}
and the log-likelihood function is thus
\begin{equation*}
\ell(\br | \bom) =  -\frac{1}{2\sigma^2}\norm{\br-\bom}_{C_1}^2 + \text{constant}
\end{equation*}
The ML estimator minimizes $\norm{\br-\bom}_{C_1}$, and is hence given by
\begin{equation}
\label{eq:projR1}
\hat\bom = \Pi_U\,\br
\end{equation}
where $\Pi_U$ denotes orthogonal projection into $U(\Gamma)$ with respect to the inner product on $C_1$. The residual $\br-\hat\bom$ then resides in the cycle space $Z(\Gamma)$.  This result provides a useful characterization of the ML estimator: the estimate satisfies Kirchhoff's voltage law and the residual satisfies Kirchhoff's current law.

To explicitly compute \eqref{eq:projR1}, it is desirable to parametrize the space $U(\Gamma)$, whose definition~(\ref{eq:cuts}) does not well elucidate the nature of its elements. Two basic parameterizations are useful. In the first, a particular vertex is chosen as a reference. Then \eqref{eq:vertexdecomp} implies that the value of $\bom\in U(\Gamma)$ is determined by the relative offsets of the other $|V(\Gamma)|-1 =  n-1$ vertices, denoted by $W$.  The second parameterization of $U(\Gamma)$ is obtained by choosing spanning tree $S\in E(\Gamma)$ of $\Gamma$ and noting that if $\bom$ is known on $S$, then all $n-1$ offsets relative to a reference vertex can be determined by following the tree.

In the first of these cases, the basis given above for $\linspan(W)$ is chosen and $\bx\in\linspan(W)$ may be expressed as
\begin{equation*}
\bx = \sum_{j=1}^{n-1} x_j \bv_{j}. 
\end{equation*}

Alternatively, for the fixed spanning tree $S$, one may write $\bnu\in \cS=\linspan(S)$ as
\begin{equation*}
\bnu = \sum_{j=1}^{n-1} \nu_j \be_{\rho(j)}
\end{equation*}
where the $\rho(j)$ label the $n-1$ edges comprising the spanning tree $S$.



These representations enable the definition of the $(n-1)\times m$ matrix $D_W$ and the $(n-1)\times (n-1)$ matrix $D_{WS}$ with entries
\begin{align*}
[D_W]_{ij} &= \langle\bv_{i}, D\be_j \rangle_{C_0}\\ 
[D_{WS}]_{ij} &= \langle\bv_{i}, D\be_{\rho(j)} \rangle_{C_0},
\end{align*}
$[D_{WS}]$ is the matrix of the restriction of $D$ to $\linspan(S)$. With these definitions,
\begin{equation}\label{eq:xtobom} 
\bom = D_W^\sT \bx
\end{equation}
and 
\begin{equation}
\label{eq:nu}
\bnu = D_{WS}^{\sT} \bx. 
\end{equation}
$D_{WS}$ is invertible; in fact \cite[p.~101]{Jungnickel2008}, 
\begin{equation}\label{eq:umod}
	\det D_{WS} = \pm 1.
\end{equation} 
Taking inverses in (\ref{eq:nu}) yields 
\begin{equation*}
\bx = \left(D_{WS}^{\sT}\right)^{-1} 
\bnu = {D_{WS}^{\sT}}^{-1} P_S\bom
\end{equation*}
where $[P_S]_{ij} = \delta_{\rho(i),j}$ is the matrix of the orthogonal projection onto $\cS$. The matrix 
\begin{equation*}
L_W = D_W D_W^\sT  
\end{equation*}
is the Laplacian (Kirchhoff) matrix with the row and column corresponding to $v_n$ removed. 

With this notation, the maximum-likelihood estimates for  $\bx$ and  $\bnu$ are then 
\begin{align}
\label{eq:ML1Dx}
\hat \bx &= L_W^{-1} D_W \br\\
\hat \bnu &= D_{WS}^\sT L_W^{-1} D_W \br  \nonumber
\end{align}
and  the estimate of $\bom$ is
\begin{equation*}
\hat \bom = D_W^{\sT} L_W^{-1} D_W \br
\end{equation*}
The Fisher information matrix for estimation of the offsets $x_j$ is
\begin{equation*}
\begin{split}
 F^W &= -\sE\left\{\bnabla^2_{\bx} \log p(\br|\bom) \right\} \\
 &= \frac{1}{\sigma^2} \left(\bnabla_{\bx}\bom\right)^\sT \bnabla_{\bx}\bom.
\end{split} 
\end{equation*}
Equation \eqref{eq:xtobom} implies that
\begin{equation}
\label{eq:dxtobom}
  \bnabla_{\bx}\bom = \left[\frac{\partial \omega_i}{\partial x_j}\right] = D_W^{\sT},
\end{equation}
and so
\begin{equation*}
F^W = \frac{1}{\sigma^2} D_W D_W^\sT=\frac{1}{\sigma^2} L_W 
\end{equation*}
Hence $F^W$ is proportional to the matrix of the Kirchhoff map in the standard basis, but with the row and column corresponding to the reference vertex removed.  In a similar way the Fisher information matrix for estimation of the $\nu_j$ is
\begin{equation*}
F^S = \frac{1}{\sigma^2} D_{WS}^\sT L_W D_{WS}.
\end{equation*}

Note that $\det L_W$ is a minor of the Kirchhoff matrix $L$, and so by the Kirchhoff
matrix tree theorem it is equal to the number of spanning trees
$t(\Gamma)$ of $\Gamma$. By \eqref{eq:umod}, the determinant of the Fisher
information is thus
\begin{equation} \label{eq:detfishR}
\begin{split}
\det F^W = \det F^S &= {\sigma^{-2(n-1)}}\det L_W\\
&=  {\sigma^{-2(n-1)}} t(\Gamma)
\end{split}
\end{equation}

The best possible situation occurs when the pairwise difference between all nodes is measured. In this case, $\Gamma$ is a complete graph and the number of spanning trees is known to be $t(\Gamma)=n^{(n-2)}$. In this situation, the ``average'' Fisher information per node is
\begin{equation*}
	(\det F^W)^{1/(n-1)} = \frac{1}{\sigma^2} n^{(n-2)/(n-1)} \sim \frac{n}{\sigma^2} \quad \text{as $n \rightarrow \infty$}
\end{equation*}

Both of the  ML estimators $\hat\bx$ and $\hat\bom$  are unbiased, and as a
result, since the covariance matrix of $\br$ is $\sigma^2 \bbI$,
\begin{equation*}
C_{\hat\bx} = \sE\{(\bx-\hat\bx)(\bx-\hat\bx)^{\sT}\} =  \frac{1}{\sigma^2} L_W,
\end{equation*}
and its determinant is 
\begin{equation}\label{eq:covariance_det}
\det C_{\hat\bx} = \frac{{\sigma^{2(n-1)}}}{t(\Gamma)},
\end{equation}
as anticipated from \eqref{eq:detfishR}. The covariance matrix of the estimator $\hat\omega$ is
\begin{equation*}
\begin{split}
C_{\hat\omega} &= \sE\{(\bom - \hat\bom)(\bom - \hat\bom)^\sT\}\\ 
&=  D_W^\sT (F^W)^{-1} D_W\\
& =  \sigma^2 D_W^\sT L_W^{-1} D_W\\
& = \sigma^2 P_{U}
\end{split}
\end{equation*}
since $P_U=D_W^\sT L_W^{-1} D_W$ is the orthogonal projection onto $U(\Gamma)$. An interesting consequence of this observation is that 
\begin{equation*}
	\tr C_{\hat\omega} = \sigma^2 \dim U(\Gamma)
\end{equation*}

\subsection{Network design for independent errors on edges}
Before going on to other measurement models, it is instructive to consider briefly the consequences of the above results in the design of a synchronization or registration scheme for a network. Since the ML estimator \eqref{eq:ML1Dx} is unbiased for any graph $\Gamma$, the role of the number of spanning trees $t(\Gamma)$ in the determinant of the estimator covariance matrix \eqref{eq:covariance_det}, or equivalently in the determinant of the Fisher information matrix \eqref{eq:detfishR}, shows that a large number of spanning trees is desirable for good estimator performance.

\begin{figure}[htb]
\centering
 \includegraphics[width=0.5\textwidth]{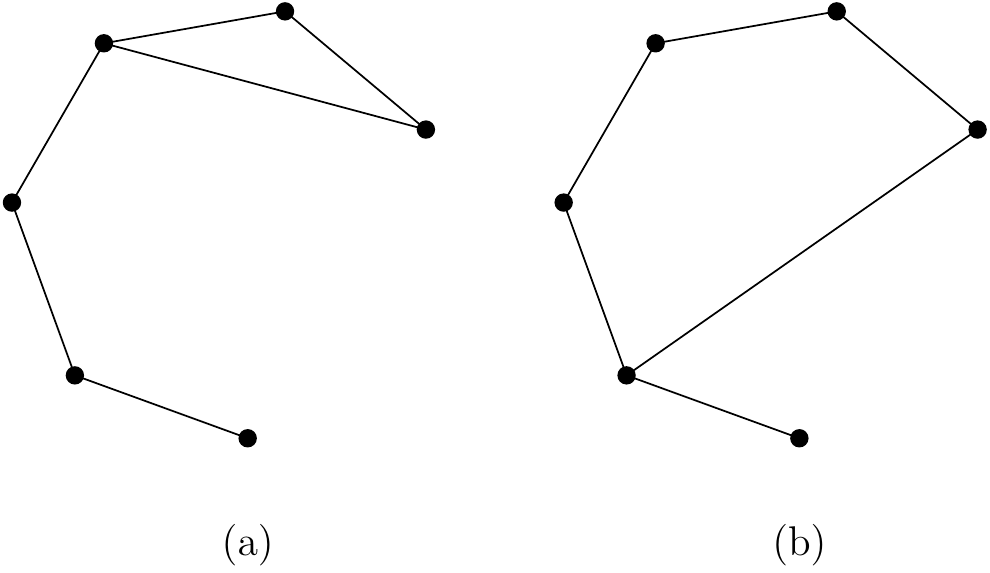}
\caption{Two networks with the same number of nodes and links. Network (a) has two
spanning trees while network (b) has four spanning trees and is hence superior in 
the estimation context of this paper.}
\label{fig:network1a}
\end{figure}

\begin{figure}[htb]
\centering
 \includegraphics[width=0.22\textwidth]{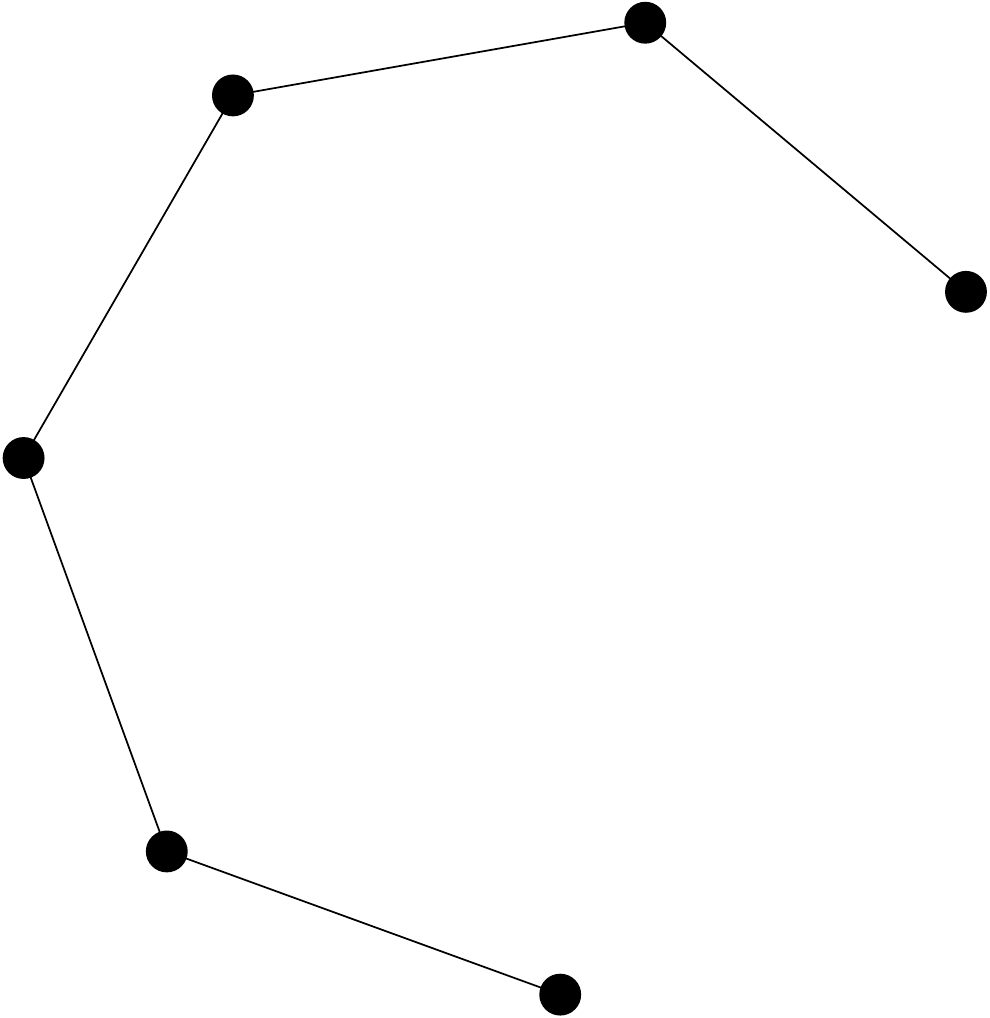}
\caption{An acyclic network.  If one link can be added, the best choice is to create a ring network. The worst choice is to create the network shown in Figure \ref{fig:network1a}(a).}
\label{fig:network2a}
\end{figure}

The network depicted in Figure \ref{fig:network1a}(a) has the same number of nodes and links as the one in Figure \ref{fig:network1a}(b).  But the number of spanning trees in the former is two while the number of spanning trees in the latter is four.  So, under the model assumed in this paper, estimation fidelity will be better for the network of Figure \ref{fig:network1a}(b). From the perspective of design, if one has the opportunity to add one link to the acyclic network shown in Figure \ref{fig:network2a}, the best choice in the context of this paper is to create a ring network (five spanning trees) and the worst is to create the network of Figure \ref{fig:network1a}(a).

\subsection{Estimation with correlated measurements}

This section further examines the situation where $G=\bbR$ and the measurement model is given by \eqref{eq:GaussR}, but now with the measurement errors, in the standard basis for $C_1(\Gamma)$, being jointly Gaussian with covariance matrix $R$. In this setting, the probability density of the measurements is
\begin{equation*}
p(\br | \bom) =  \frac{1}{\sqrt{(2\pi)^m\det R}}\exp \left(-\frac{1}{2} (\br - \bom)R^{-1}(\br - \bom)^\sT\right)
\end{equation*}
The maximum-likelihood estimate of $\bom\in U(\Gamma)$ is obtained by splitting the data $\br$ as
\begin{equation}
\label{eq:kircest}
\br = \hat\bom + \hat\beps,
\end{equation}
where $\hat\bom\in U(\Gamma)$ and residual $\hat\beps$ satisfies $R^{-1}\hat\beps\in Z(\Gamma)$. That is,
\begin{equation*}
\hat\bom = Q_U \br
\end{equation*}
where $Q_U$ is now an oblique projection with range $U(\Gamma)$ and null space $R^{-1} (Z(\Gamma))$.
\begin{figure}[h]
\includegraphics[width=\linewidth]{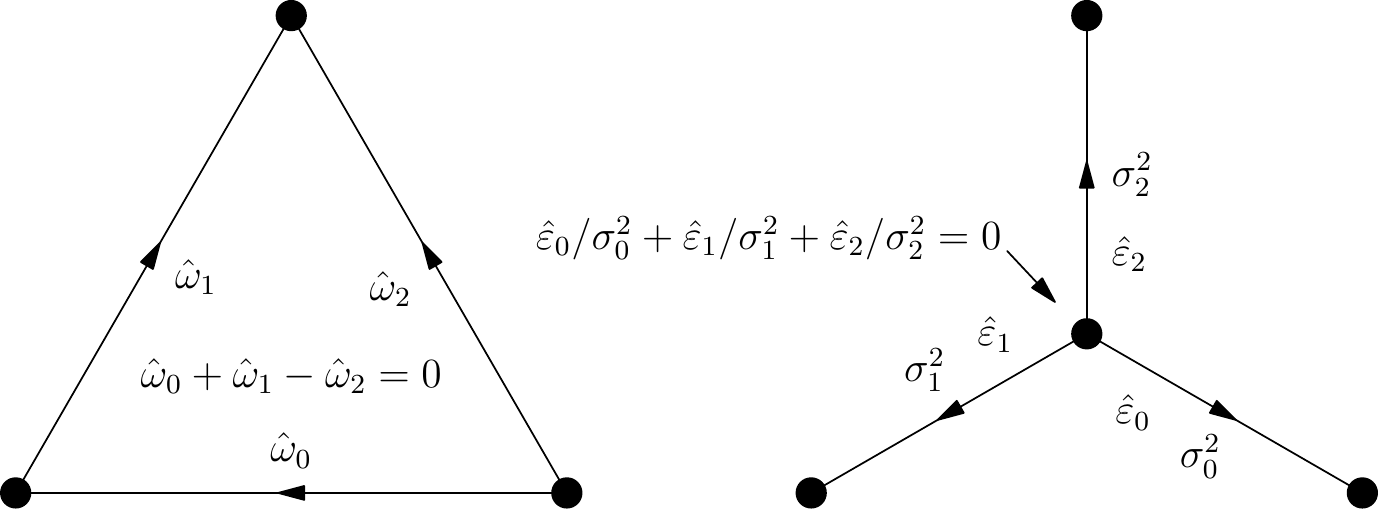}
\caption{The residual $\beps$ of the ML estimate has the $R^{-1}\hat\beps$ satisfies the Kirchhoff current law.}
\label{fig:wkirchhoff}
\end{figure}
This situation is illustrated in Figure~\ref{fig:wkirchhoff} for a diagonal covariance matrix with entries $\sigma^2_j$. The estimate $\hat\bom$ satisfies the Kirchhoff voltage law; i.e., the oriented sum of $\hat\bom$ around any cycle as intended. It is interesting to observe that it is now $R^{-1}\hat\beps$ that satisfies the Kirchhoff current law; i.e., the oriented sum at any vertex is zero, with the covariance $R$ playing the role of resistance in a way akin to Ohm's law. 

The columns of $D_W^\sT$ are a basis for $U(\Gamma)$ and \eqref{eq:1} implies that the columns of $R^{-1}D_W^\sT$ are a basis for $(R^{-1}(Z(\Gamma)))^\perp$. Thus, 
\begin{equation*}
\hat\bom = D_W^\sT \left( D_W R^{-1} D_W^\sT \right)^{-1}  D_W R^{-1}\br
\end{equation*}
The corresponding ML estimate for the vertex offsets $\bx$ is
\begin{equation*}
\hat\bx = \left( D_W R^{-1} D_W^\sT \right)^{-1} D_W R^{-1}\br
\end{equation*} 
Motivated by these expressions, it is convenient to define the weighted Laplacian 
\begin{equation*}
\tilde L = D R^{-1} D^\sT
\end{equation*}
and similarly
\begin{equation*}
\tilde L_W = D_W R^{-1} D_W^\sT
\end{equation*}

By \eqref{eq:dxtobom}, the Fisher information matrix for the vertex parametrization is given by
\begin{equation*}
\begin{split}
F^W &= -\sE\left\{\bnabla^2_{\bx}  \log p(\br|\bom) \right\} \\
&= D_W R^{-1} D_W^\sT = \tilde L_W 
\end{split} 
\end{equation*}

The Cauchy-Binet formula \cite{Horn1993} allows the determinant of the Fisher information to be written as
\begin{equation*}
\begin{split}
 \det F^W &= \det (D_W R^{-1} D_W^\sT)\\
& = \sum_{T}\sum_{T'} \det (D_{WT})  \det([R^{-1}]_{TT'}) \det (D_{WT'})
\end{split}
\end{equation*}
where $T$ and $T'$ denote subsets of columns (edges) which are retained. The sum extends over all subsets of $E(\Gamma)$ of order $n-1$. The matrices $D_{WT}$ have the property \cite[p.~101]{Jungnickel2008}
\begin{equation*}
\det D_{WT} = \begin{cases}\pm 1, & \text{if $T$ is a spanning tree,}\\ 0, &\text{otherwise.}\end{cases}
\end{equation*} 
The only terms in the above sum over $T$ which are non-zero are those corresponding to spanning trees and
so
 \begin{equation}\label{eq:corrFish}
 \det F^W =  \sum_{S,S} \alpha_{SS'}\; \det ([R^{-1}]_{SS'})
\end{equation}
The quantity 
\begin{equation}\label{eq:alpha}
\alpha_{SS'}=\det( D_{WS}D_{WS'}^\sT)
\end{equation} 
takes values $\pm 1$ depending on the pair of spanning trees $S$ and $S'$ and independently of the choice of $W$. 

With the assumption $R=\diag(\sigma_1^2,...,\sigma_m^2)$, \eqref{eq:corrFish} simplifies
to
\begin{equation*}
  \det F^W = \sum_{S} \prod_{e_j\in S} \frac{1}{\sigma^2_j},
\end{equation*}
which reduces to further to \eqref{eq:detfishR} when $\sigma^2_j = \sigma^2$ for all $e_j\in E(\Gamma)$.

The ML estimate of the vertex offsets $\bx$ is unbiased and its covariance matrix is 
\begin{equation*}
C_{\hat\bx} = \sE\{(\bx-\hat\bx)(\bx-\hat\bx)^\sT\} = \tilde L_W^{-1}
\end{equation*}
and has determinant
\begin{equation*}
\det C_{\hat\bx} = \frac{1}{\sum_{S,S} \alpha_{SS'}\; \det ([R^{-1}]_{SS'})}
\end{equation*}
from \eqref{eq:corrFish}. The ML estimate of $\bom$ is also unbiased and has covariance
\begin{equation*}
\begin{split}
C_{\hat\bom} = \sE\{(\bom-\hat\bom)(\bom-\hat\bom)^\sT\} &= D_W ^\sT\tilde L_W^{-1} D_W\\
				   &= Q_U R
\end{split}
\end{equation*}

\subsection{Multi-dimensional Gaussian problem}
\label{sec:Rdgauss}
This section further generalizes the setting to $G=\bbR^d$, where the state of each vertex in the network is a vector in $\bbR^d$. In this situation, it is necessary to consider more general functions of the graph $\Gamma$ than were treated in Section \ref{sec:graphs}. The vertex space now consists of functions $V(\Gamma)\rightarrow \bbR^d$ and is correspondingly denoted by $C_0(\Gamma,\bbR^d)$. In fact,
\begin{equation*}
C_0(\Gamma, \bbR^d) = {\bbR}^d \otimes C_0(\Gamma, \bbR)
\end{equation*}
In terms of the standard basis $\bq_j$, $j=1,\cdots, d$ for $\bbR^d$, any element of $C_0(\Gamma, \bbR^d)$ can be expressed as
\begin{equation*}
\bx = \sum_{i=1}^d \sum_{j=1}^n x_{ij} \bq_i \otimes \bv_j
    = \sum_{j=1}^n \bx_j\otimes \bv_j
\end{equation*}
Similarly, the vector space of functions $E(\Gamma)\rightarrow \bbR^d$ is
\begin{equation*}
C_1(\Gamma,\bbR^d) = \bbR^d \otimes C_1(\Gamma,\bbR)
\end{equation*}
The boundary map on $C_1(\Gamma,\bbR^d)$ is $\bbI\otimes D$ where $\bbI$ is the identity map on $\bbR^d$ and $D$ is the boundary map on $C_1(\Gamma)$. The coboundary map on $C_0(\Gamma,\bbR^d)$ is $\bbI\otimes D^\sT$.

\begin{figure}[h]
\begin{center}
\begin{asy}
usepackage("amsmath");
usepackage("amssymb");
size(4cm);
draw(Label("$\bbI \otimes D$", MidPoint, N), (0,0.3){NE}..{SE}(6,0.3), Arrow);
label("$C_1(\Gamma, \bbR^d)$",(0,0),W);
draw(Label("$\bbI\otimes D^{\sT}$", MidPoint, S), (6,-0.3){SW}..{NW}(0,-0.3), Arrow);
label("$C_0(\Gamma, \bbR^d)$",(6,0),E);
\end{asy}
\end{center}
\end{figure}

As in the one-dimensional case, the cycle space $Z(\Gamma,\bbR^d)$ is defined to be the kernel of the boundary map; i.e., the set of $\bz \in C_1(\Gamma, \bbR^d)$ such that  
\begin{equation*}
(\bbI\otimes D) \bz = 0.
\end{equation*}
Writing
\begin{equation*}
    \bz = \sum_{j=1}^n \bz_j\otimes \be_j
\end{equation*}
the cycles satisfy a vector form of \eqref{eq:kirchhoff}.
\begin{equation*}
\sum_{e_j: t(e_j)=v} \bz_j = \sum_{e_j: s(e_j)=v} \bz_j,
\end{equation*}
i.e., at any vertex, the oriented vector sum of the values of $\bz$ on the set of edges meeting at the vertex is zero.
 
The cocycle space $U(\Gamma, \bbR^d)$ is the image of the coboundary map $\bbI\otimes D^\sT$. For any 
\begin{equation*}
\bom = \sum_{j=1}^n \bom_j\otimes \be_j \in U(\Gamma, \bbR^d)
\end{equation*}
the equation
\begin{equation*}
	\sum_{e_j\in \fL} (-1)^{\sigma_j} \bom_j = 0,
\end{equation*}
must hold for for all cycles $\fL$.  In this expression, $\sigma_j = 0$ if $e_j$ is oriented as $\fL$ and $\sigma_j = 1$ if $e_j$ is oriented opposite to $\fL$.

An inner product can be defined on $C_1(\Gamma,\bbR^d)$, based on the inner
product defined earlier on $C_1(\Gamma)$ and the standard inner product on
$\bbR^d$, namely, for $\bs_1, \bs_2\in\bbR^d$ and $\bx_1, \bx_2\in
C_1(\Gamma)$,
\begin{equation*}
\langle \bs_1\otimes\bx_1, \bs_1\otimes\bx_1\rangle = \langle \bs_1, \bs_2\rangle_{\bbR^d} \langle \bx_1, \bx_2\rangle_{C_1(\Gamma)}
\end{equation*}
For all $\bz\in Z(\Gamma, \bbR^d)$ and $\bom\in U(\Gamma, \bbR^d)$
\begin{equation*}
\langle \bz, \bom\rangle = 0,
\end{equation*}
i.e., $Z(\Gamma,\bbR^d)$ is the orthogonal complement of $U(\Gamma,\bbR^d)$.

\subsubsection{Independent identically distributed edge measurements}

Recall that the  measurement model is $\br = \bom + \beps$ with
$\beps\sim\cN(\bzero,R)$, $\br$ and $\beps$ in $C_1(\Gamma,\bbR^d)$, and
$\bom\in U(\Gamma,\bbR^d)$. The first case of interest is when the errors
$\beps_j\in\bbR^d$ are independent.  Their individual probability densities
are
\begin{equation*}
 p(\beps_j) = \frac{1}{\sqrt{(2\pi)^d\det R}} \exp\left( -\frac{1}{2}\langle\beps_j,  R^{-1} \beps_j\rangle_{\bbR^d} \right)
\end{equation*}
for $j = 1, \cdots, m$. By independence of the $\beps_j$, the joint probability density for the data $\br\in C_1(\Gamma,\bbR^d)$ is thus
\begin{equation*}
 p(\br | \bom) = \frac{1}{((2\pi)^d\det R)^{m/2}}\exp -\frac{1}{2} \langle \br-\bom, (R^{-1}\otimes I) (\br-\bom)\rangle
\end{equation*}
The maximum likelihood estimate of $\bom$ is obtained by splitting the data $\br$ as
\begin{equation*}
\br = \hat \bom + \hat \beps,
\end{equation*}
where $\hat\bom\in U(\Gamma,\bbR^d)$ and $\hat\beps\in Z(\Gamma,\bbR^d)$. Explicitly, $\omega\in U(\Gamma,\bbR^d)$ is parameterized in terms of the offsets of $n-1$ vertices with respect to a reference vertex
\begin{equation*}
\bom = (\bbI\otimes D_W^\sT)\bx
\end{equation*}
or alternatively, in terms of a spanning tree $S$, as
\begin{equation}\label{eq:vxtobom}
\bom = (\bbI\otimes D_W^\sT D_{WS}^{-1})\bnu
\end{equation}
where $\bx$ and $\bnu$ are related by
\begin{equation*}
\bx = (\bbI\otimes D_{WS}^{-1})\bnu
\end{equation*}
In terms of these parameterizations, the maximum-likelihood estimate is
\begin{equation*}
\begin{split}
\hat\bx &= (R^{-1}\otimes L_W^{-1} )(R^{-1}\otimes D_W)\br\\
&=(\bbI\otimes L_W^{-1}D_W) \br,
\end{split}
\end{equation*}
and 
\begin{equation*}
\hat\bnu = (\bbI\otimes D_{WS}^{-1}L_W^{-1}D_W) \br
\end{equation*}
So $\hat\bom = (\bbI\otimes D_W^\sT L_W^{-1}D_W) \br$.

The Fisher information matrix for $\bx$ is given by
\begin{equation*} 
\begin{split}
 F^W &= -\sE\left\{\bnabla^2_{\bx} \log p(\br | \bom) \right\} \\
&= \left(\bnabla_{\bx}\bom\right)^\sT (\bbI\otimes R^{-1})\bnabla_{\bx}\bom
\end{split} 
\end{equation*}
Equation \eqref{eq:vxtobom} implies that
\begin{equation*}
\bnabla_{\bx}\bom = \bbI\otimes D_W^\sT,
\end{equation*}
so that
\begin{equation*}
\begin{split}
F^W &= R^{-1}\otimes D_W D_W^\sT\\ 
&=R^{-1}\otimes L_W 
\end{split} 
\end{equation*}
In a similar way, the Fisher information matrix for estimation of $\bnu$ is
\begin{equation*}
F^S = R^{-1}\otimes D_{WS}^\sT L_W D_{WS}.
\end{equation*}
The determinant of the Fisher information in both cases is
\begin{equation*}
\det F^W = \det F^S =\frac{(\det L_W)^d}{(\det R)^{n-1}}
= \frac{t(\Gamma)^d}{(\det R)^{n-1}},
\end{equation*}
which is obtained by using the tensor product identity
\begin{equation*}
\det A\otimes B = (\det A)^{\dim B} (\det B)^{\dim A}.
\end{equation*}

Finally, the error covariance of the unbiased ML estimate $\hat\bom$ is
\begin{equation*}
\begin{split}
C_{\hat\bom} &= \sE\{(\bom-\hat\bom)(\bom-\hat\bom)^\sT\}\\ &=  R\otimes D_W^\sT L_W^{-1} D_W = R\otimes P_U,
\end{split}
\end{equation*}
where $P_U$ is the orthogonal projection onto $U(\Gamma)$. Taking the partial trace over 
$C_1(\Gamma)$ yields
\begin{equation*}
\tr_{C_1(\Gamma)} C_{\hat\bom}  = (\dim U(\Gamma)) R.
\end{equation*}

\subsubsection{Correlated edge measurements}
Now assume the errors on the edge measurements $\beps_j\in\bbR^d$ have joint probability density
\begin{equation*}
p(\beps) = \frac{1}{\sqrt{(2\pi)^{md}\det R}} \exp\left( -\frac{1}{2}\langle\beps,  R^{-1} \beps\rangle_{C_1(\Gamma,\bbR^d)} \right)
\end{equation*}
where $R$ now denotes the $md\times md$ covariance matrix of the $m$ $d$-dimensional edge measurements. The probability density for the data $\br\in C_1(\Gamma,\bbR^d)$ is
\begin{equation*}
p(\br | \bom) = 
\frac{1}{\sqrt{(2\pi)^{md}\det R}} \exp\left( -\frac{1}{2}\langle\br-\bom,  R^{-1} \br-\bom\rangle \right)
\end{equation*}
where the inner product in the exponent is in $C_1(\Gamma,\bbR^d)$. The ML
estimate involves a decomposition of the data as
\begin{equation*}
\br = \hat \bom + \hat \beps,
\end{equation*}
where $\hat\bom\in U(\Gamma,\bbR^d)$ and $\hat\beps\in R^{-1}(Z(\Gamma,\bbR^d))$.
Explicitly, 
\begin{equation*}
\hat\bom = Q_U \br,
\end{equation*}
where $Q_U$ is the oblique projection with range $U(\Gamma,\bbR^d)$ and null space $R^{-1}(Z(\Gamma, \bbR^d))$; i.e.,
\begin{equation*}
Q_U = (\bbI\otimes D_W^\sT) \left((\bbI\otimes D_W) R^{-1} (\bbI\otimes D_W^\sT)\right)^{-1} (\bbI\otimes D_W) R^{-1}
\end{equation*}
The ML estimate of the vertex offsets is
\begin{equation*}
\hat\bx = \left((\bbI\otimes D_W)R^{-1} (\bbI\otimes D_W^\sT)\right)^{-1} (\bbI\otimes D_W)R^{-1}\br.
\end{equation*}

The Fisher information for $\bx$ is
\begin{equation*}
F^W = (\bbI\otimes D_W)R^{-1}(\bbI\otimes D_W^\sT)
\end{equation*}
In order to calculate the determinant of $F^W$, it is helpful to first find 
\begin{equation*}
\det(\bbI\otimes D_W)_\cT
\end{equation*}
where $\cT$ denotes a subset of $(n-1)d$ of the columns. Now $(\bbI\otimes D_W)$ has the form
\begin{equation*}
\begin{pmatrix} D_W & \bzero & \bzero & \bzero\\ \bzero & D_W & \bzero & \bzero \\ \bzero & \bzero & \ddots & \bzero\\ \bzero & \bzero & \bzero & D_W\end{pmatrix}
\end{equation*}
Write $\cT = (T_1, T_2, \cdots, T_d)$, where $T_k$ denotes the set of columns selected from the $k^{\text{th}}$
block. Since $\Gamma$ is assumed to be connected, $D_W$ has rank $n-1$ thus $\det (\bbI\otimes D_W)_{\cT}=0$ unless exactly $n-1$ columns are chosen from each block, in which case
\begin{equation*}
\begin{split}
\det(\bbI\otimes D_W)_{\cT} &= \prod_{j=1}^d \det(D_{WT_j})\\
& = \begin{cases} \pm 1, & \text{if each $T_j$ is a spanning tree} \\ 0, & \text{otherwise,}\end{cases}
\end{split}
\end{equation*}
$\cS = (S_1,S_2,\ldots,S_d)$ will be called a multi-spanning tree if it consists of one spanning tree for each dimension of the parameter space $R^d$. Applying the Cauchy-Binet formula gives
\begin{equation}
\label{eq:Rdfisher}
\begin{split}
\det F^W &= \sum_{\cT\cT'} \det(\bbI\otimes D_W)_{\cT} \det([R^{-1}]_{\cT\cT'})\det (\bbI\otimes D_W^{\sT})_{\cT} \\
&=  \sum_{\cS\cS'} \alpha_{\cS\cS'}\det([R^{-1}]_{\cS\cS'}) 
\end{split} 
\end{equation} 
where the sum is over multi-spanning trees.  In this expression,
\begin{equation*}
\alpha_{\cS\cS'} = \prod_{j=1}^d \alpha_{S_j S'_j}
\end{equation*}
with $\alpha_{SS'}$ given by \eqref{eq:alpha}, which takes values $\pm 1$. If the edge measurement errors
are independent then \eqref{eq:Rdfisher} reduces to
\begin{equation}
\label{eq:Rdfisherind}
\det F^W = \sum_{\cS} \det([R^{-1}]_{\cS\cS}). 
\end{equation} 

The error covariance of the ML estimate of the vertex offsets $\bx$ is
\begin{equation*}
\sE\{(\bx-\hat\bx)(\bx-\hat\bx)^\sT\} = \left((\bbI\otimes D_W) R^{-1} (\bbI\otimes D_W^\sT)\right)^{-1}
\end{equation*}
So,
\begin{equation*}
\det \sE\{(\bx-\hat\bx)(\bx-\hat\bx)^\sT\} = \frac{1}{\sum_{\cS \cS'} \alpha_{\cS \cS'}\det([R^{-1}]_{\cS \cS'}) }
\end{equation*}
and
\begin{equation*}
\begin{split}
&\sE\{(\bom-\hat\bom)(\bom-\hat\bom)^\sT\}\\
 &= (\bbI\otimes D_W^\sT) \left((\bbI\otimes D_W)R^{-1}(\bbI\otimes D_W^\sT)\right)^{-1}(\bbI\otimes D_W^\sT)\\
				   &= Q_U R
\end{split}
\end{equation*}

\section{Local Estimators for Gaussian Noise on $\bbR$}
\label{sec:localRd}

The estimators presented thus far have either implicitly or explicitly assumed
a fixed reference vertex or a particular spanning tree. This is intrinsically
incompatible with the development of local estimators; i.e., estimators that
can be implemented in ways that each vertex follows some procedure that uses
only information 
accessible from  its nearest neighbors. Consider again the situation involving
Gaussian noise on $\bbR$ with the noise on edge measurements being
independent, treated in Section~\ref{sec:GaussRind}, and recall that the ML
estimate for $\bom$ is such that the residual $(\br-\hat\bom)\in Z(\Gamma)$;
i.e., it obeys Kirchhoff's current law. Thus, $\hat\bom$ is the solution of
\begin{equation*}
D \hat\bom = D\br.
\end{equation*} 
Indeed, if $\bx$ is any element of $C_0(\Gamma)$ satisfying 
\begin{equation*}
\label{eq:localML1}
L\bx = D\br
\end{equation*}
then $\hat\bom = D^\sT\bx$ will be the ML estimate of $\bom$. The quantity
$\bx$ may be interpreted as the collection of offsets each vertex needs to
apply to its own coordinate in order for the entire network to be aligned in a
statistically optimal way. It is interesting to note that addition of a
positive multiple of the orthogonal projection onto the subspace of constant
functions on $V(\Gamma)$ to the Laplacian matrix creates an invertible
matrix. Hence a solution to \eqref{eq:localML1} is obtained from
\begin{equation*}
\bx = (L+\frac{\mu}{n} \bone\bone^\sT)^{-1} D\br
\end{equation*}
where $\bone$ denotes a vector of ones. Since $\bone^\sT D = 0$, this is the same solution for all $\mu > 0$. In fact, it is the solution satisfying $\bone^\sT \bx = 0$; i.e., the set of offsets obtained do not change the mean of the vertex states across the network.

If the linear system \eqref{eq:localML1} is solved using Jacobi's method
\cite{Golub1996}, the structure of the Laplacian matrix $L$ ensures this
algorithm is local. Jacobi's method involves writing $L=N-A$ in terms of the
diagonal degree matrix $N$ and the adjacency matrix $A$ and applying the
recursion
\begin{equation}\label{eq:jacobiRec}
\bx^{(t+1)} = N^{-1} \left( D\br + A \bx^{(t)}\right)
\end{equation}
A fixed point of this recursion satisfies \eqref{eq:localML}. Thus if the method converges, it gives the ML estimate. Jacobi's method is known to converge if the matrix $L$ is diagonally dominant \cite{Weiss2001}; i.e., 
\begin{equation}\label{jacobiConv}
\abs{L_{ii}} > \sum_{j\neq i} \abs{L_{ij}}, 
\end{equation}
although this is not a necessary condition. The Laplacian matrix of a graph satisfies
\begin{equation*}
\abs{L_{ii}} = \sum_{j\neq i} \abs{L_{ij}}.
\end{equation*}
Another sufficient condition for convergence of Jacobi's method is the so-called \emph{walk-summability} condition \cite{Malioutov2009}. This specifies that the spectral radius of $N^{-1}A$ is less than one; i.e.,
\begin{equation}\label{eq:specrad}
\rho(N^{-1}A) < 1.
\end{equation}
The Gershgorin circle theorem \cite{Horn1993} implies that $\rho(N^{-1}A)\leq 1$.

The Jacobi algorithm has been applied in MAP estimation using Gaussian belief propagation in Bayesian belief networks \cite{Johnson2009}. In this work, it is noted that when  neither \eqref{jacobiConv} and \eqref{eq:specrad} are satisfied and Jacobi's method fails to converge, convergence can forced by using a double-loop iterative method. In the estimation problem of interest in this section, the possible violations of the sufficient conditions for convergence are as mild as can be found in practice. In experiments run to date, Jacobi's method has never failed to converge. 

Once the recursion \eqref{eq:jacobiRec} is written out in detail, the update
for the $k^{\text{th}}$ vertex becomes
\begin{equation*}
x^{(t+1)}_k = \frac{1}{n_k} \sum_{v_\ell \sim v_k}\left( x^{(t)}_\ell + r_{(\ell,k)}\right),
\end{equation*}
where $r_{(\ell,k)}$ is $r_e$ if $e=(v_\ell,v_k)\in E(\Gamma)$ or $-r_e$ if
$e=(v_k,v_\ell)\in E(\Gamma)$. Thus the recursion is indeed local. Note that
$x^{(t)}_\ell+r_{(\ell,k)}$ is the current prediction at the neighboring
vertex $\ell$ of the value at the $k^\text{th}$ should be. At each
vertex, a single iteration of the algorithm can be summarized as ``become the
mean of what your neighbors say your value should be.''

An alternate way to state the recursion, which will be important subsequent application to estimation in Lie groups other that $\bbR^d$, is as follows. Denote an action of $\bbR$ on itself by
\begin{equation*}
	T_r x = x + r
\end{equation*}
Then \eqref{eq:jacobiRec} can be written as
\begin{equation*}
\bx^{(t+1)} = Q \bx^{(t)} 
\end{equation*}
where
\begin{equation*}
Q_{k,\ell} = \begin{cases} \frac{1}{n_k} T_{r_{(k,\ell)}},& \text{if $(v_\ell,v_k) \in E(\Gamma)$},\\
\frac{1}{n_k} T_{r_{(k,\ell)}}^{-1},& \text{if $(v_k,v_\ell) \in E(\Gamma)$},\\ 
0,& \text{otherwise}.\end{cases}
\end{equation*}

Although it may be possible to develop local algorithms for some forms of correlation between measurement noise on different edges of the graph, this possibility is not explored here. For $G=\bbR^d$, any $\bx\in C_1(\Gamma,\bbR^d)$ satisfying 
\begin{equation}\label{eq:localMLRd}
(\bbI\otimes L)\bx = (\bbI\otimes D)\br
\end{equation}
then $\hat\bom = (\bbI\otimes D^\sT)\bx$ will be the ML estimate of $\bom$. Jacobi's method in this case give the recursion
\begin{equation}\label{eq:jacobiRecRd}
\bx^{(t+1)} = (\bbI\otimes N)^{-1} \left( (\bbI\otimes D)\br + (\bbI\otimes A) \bx^{(t)}\right)
\end{equation}
The fixed points of \eqref{eq:jacobiRecRd} satisfy \eqref{eq:localMLRd} and thus 
\begin{equation*}
\abs{(\bbI\otimes L)_{ii}} = \sum_{j\neq i} \abs{(\bbI\otimes L)_{ij}}.
\end{equation*}
and by the Gershgorin circle theorem
\begin{equation*}
\rho((\bbI\otimes N)^{-1}(\bbI\otimes A)) \leq 1.
\end{equation*}
In terms of the action of $\bbR^d$ on itself
\begin{equation*}
	T_\br = \bx + \br,
\end{equation*}
\eqref{eq:jacobiRecRd} becomes
\begin{equation*}
	\bx^{(t+1)} = Q \bx^{(t)} 
\end{equation*}
where the operator $Q$ has a ``block'' form with the $(k,\ell)^{\text{th}}$ block being
\begin{equation*}
Q_{k,\ell} = \begin{cases} \frac{1}{n_k} T_{\br_{(k,\ell)}},& \text{if $(v_\ell,v_k) \in E(\Gamma)$},\\
\frac{1}{n_k} T_{\br_{(k,\ell)}}^{-1},& \text{if $(v_k,v_\ell) \in E(\Gamma)$},\\ 
0,& \text{otherwise}.\end{cases}
\end{equation*}

\section{Phase Alignment}
\label{sec:circle}

This section addresses the case of phase estimation in a network. For this
problem, the natural parameter space at the vertices  is the group of real numbers
modulo $1$, or equivalently the circle group comprised of complex numbers with
absolute value one; i.e., $\bbT=\{e^{2\pi i\theta}:\theta \in [0,1)\}$. In
what follows, it is convenient to use the latter description.

In the preceding sections, the offsets between measurements at vertices
naturally form elements of a vector space.  In the current situation this will
no longer be true, resulting in significant differences in both theory and
algorithms. These issues are elucidated here in the context of measurements on
the circle $\bbT$ and in slightly more generality in Section
\ref{sec:abelian_Lie}, where the data reside in a connected abelian Lie group.
The setting of a compact (not necessarily abelian) Lie group, which includes
the important practical situation $G=SO(3)$, will be addressed in a sequel to
this paper.

\subsection{Cycles and Cocycles}

Following the treatment in Section \ref{sec:graphs}, denote by
$C_0(\Gamma,\bbC)$ the collection of functions from the vertices $V(\Gamma)$
to $\bbC$. Similarly, denote by $C_1(\Gamma,\bbC)$ the vector space of
complex-valued functions on the edges $E(\Gamma)$. Any element of
$C_0(\Gamma,\bbC)$  can be written in terms of the basis functions \eqref{eq:sbv} as
\begin{equation*}
  \bx = \sum_{j=1}^n x_j \bv_j
\end{equation*}
where the $x_j$ are now complex numbers. Similarly, the basis functions \eqref{eq:sbe} can be used to write any element of $C_1(\Gamma,\bbC)$ as
\begin{equation*}
  \bx = \sum_{j=1}^n x_j \be_j
\end{equation*}
with $x_j\in \bbC$. The definitions of the incidence map, etc., then generalize in a straightforward way to this situation. 
 
In this setting, the appropriate space for the measurement data is the
collection of functions from the edges $E(\Gamma)$ to $\bbT$. Motivated by
analogy with the preceding cases, this will be denoted by $C_1(\Gamma,\bbT)$.
Also as earlier, $C_0(\Gamma,\bbT)$ will denote the collection of functions
from $V(\Gamma)$ to $\bbT$. Both $C_1(\Gamma,\bbT)$ and $C_0(\Gamma,\bbT)$ are
groups under pointwise multiplication of complex numbers. However, although
$C_0(\Gamma,\bbT)\subset C_0(\Gamma,\bbC)$ and $C_1(\Gamma,\bbT)\subset
C_1(\Gamma,\bbC)$, neither $C_0(\Gamma,\bbT)$ or $C_1(\Gamma,\bbT)$ are vector
spaces. In the absence of a vector space structure, in particular the
abilities to use bases and to use inner products to identify spaces with their
duals, it is necessary to adopt a modified approach to the definition of  the boundary and
coboundary operators. This begins by considering the ``spaces'' of functions
$C_0(\Gamma,\bbZ)\subset C_0(\Gamma,\bbC)$ from  the set of vertices of
$\Gamma$ to the integers $\bbZ$ and $C_1(\Gamma,\bbZ)\subset C_1(\Gamma,\bbC)$
from $E(\Gamma)$ to $\bbZ$. An element of $C_0(\Gamma,\bbZ)$ can be expressed
in terms of the basis functions for $C_0(\Gamma,\bbC)$, which are themselves
elements of $C_0(\Gamma,\bbZ)$, as
\begin{equation*}
  \bz = \sum_{j=1}^n m_j \bv_j, \quad m_j\in \bbZ
\end{equation*}
It is straightforward to verify that set  $C_0(\Gamma,\bbZ)$ is an abelian group.  Similarly, an element of $C_1(\Gamma,\bbZ)$ can be expanded in terms of the basis elements $\be_j$ of $C_1(\Gamma,\bbC)$ with integer coefficients.


In place of the inner product are ``pairings'' between  $C_i(\Gamma,\bbZ)$ and $C_i(\Gamma,\bbT)$ given by 
\begin{equation}
  \label{eq:16}
  \begin{aligned}[t]
  \langle \bz,\btau\rangle_{\bbT,1}&=\prod_{e\in E(\Gamma)}\tau_e^{z_e}\quad (\bz\in C_{1}(\Gamma,\bbZ),\  \btau\in C_{1}(\Gamma,\bbT)),\\    
  \langle \bz,\btau\rangle_{\bbT,0}&=\prod_{v\in V(\Gamma)}\tau_v^{z_v}\quad (\bz\in C_{0}(\Gamma,\bbZ),\  \btau\in C_{0}(\Gamma,\bbT)), 
  \end{aligned}
\end{equation}
where, in each case, $\tau^{-1}=\overline{\tau}$ is the complex conjugate of $\tau\in\bbT$. 

The boundary operator $D:C_1(\Gamma,\bbC)\to C_0(\Gamma,\bbC)$ is defined, as in Section \ref{sec:graphs} on the ``basis'' elements by
\begin{equation}
  \label{eq:14}
  D(\be)=t(e)-s(e) \quad (\be \in C_{1}(\Gamma,\bbC)),
\end{equation}
with the convention that a vertex is identified with the corresponding member of the basis of $C_0(\Gamma,\bbC)$. When restricted to $C_1(\Gamma,\bbZ)$, the operator $D$ maps $C_1(\Gamma,\bbZ)$ into $C_0(\Gamma,\bbZ)$. Similarly, $D^\sT$ maps $C_0(\Gamma,\bbZ)$ into $C_1(\Gamma,\bbZ)$ and the Laplacian $L=DD^\sT$, adjacency, and degree maps all map $C_0(\Gamma,\bbZ)$ into $C_0(\Gamma,\bbZ)$.



The functions $\bz_{\mathfrak L}$ defined in \eqref{eq:cycle1}, where $\mathfrak L$ is taken over all cycles, are functions in $C_1(\Gamma,\bbZ)$. The cycle space $Z(\Gamma, \bbZ)\subset C_1(\Gamma,\bbZ)$ is now regarded as a set of integer combinations of the $\bz_{\mathfrak L}$; i.e., it consists of integer sums $\sum_{k}n_{k}\bz_{\mathfrak{L_{k}}}$ where $\mathfrak L_{k}$ are cycles. The cycle space  is the kernel of the boundary operator \eqref{eq:14} acting on $C_1(\Gamma,\bbZ)$, so $Dz = 0$ for all $\bz\in C_1(\Gamma,\bbZ)$. Note that $Z(\Gamma,\bbZ)\subset Z(\Gamma,\bbC)$.



With $\bx:V(G)\to \bbT \in C_{0}(\Gamma,\bbT)$, the $\bbT$-coboundary operator can be defined as 
\begin{equation*}
  \label{eq:11}
  D_{*}(\btau)(e)= \overline{\tau}_{t(e)}\tau_{s(e)}.
\end{equation*}
This is the dual operator of $D$ using the pairings $\langle\cdot,\cdot\rangle_{\bbT,i}$  defined in \eqref{eq:16}; i.e.,
\begin{equation*}
  \label{eq:15}
  \langle \bz,D_{*}\btau\rangle_{\bbT,0}=  \langle D\bz,\btau\rangle_{{\bbT,1}},
\end{equation*}
for all $\bz\in C_0(\Gamma,\bbZ)$ and $\btau\in C_1(\Gamma,\bbT)$.
The $\bbT$-cocycle space is defined to be the image of the operator $D_{*}$:
\begin{equation*}
  \label{eq:12}
  U(\Gamma,\bbT)=D_{*}(C_{0}(\Gamma,\bbT))\subset C_{1}(\Gamma,\bbT).
\end{equation*}
Members $\bom$ of $U(\Gamma,\bbT)$ share a property corresponding to~\eqref{eq:cocycle1}; i.e., 
\begin{equation*}
  \label{eq:13}
  \langle \bz,\bom\rangle_{{\bbT,1}} = \prod_{e\in E(\Gamma)}\omega_e^{z_e}=1
\end{equation*}
for all $\bz\in Z(\Gamma,\bbZ)$. In other words, they are ``orthogonal'' to the cycle space $Z(\Gamma,\bbZ)$ in the sense of the pairing defined in \eqref{eq:16}. This is the analogue of Kirchhoff's voltage law in this setting: the oriented product of the elements of $\bom$ around any cycle is the identity element of $\bbT$ (i.e., $1$). 


\subsection{Distributions}
\label{sec:circ_dist}

A crucial issue in the analysis of the phase alignment problem is specification
of an appropriate model for noise on $\bbT$. Two distributions are typically
considered for circular statistics \cite{Mardia2001}: the wrapped normal and
the von Mises distributions. While the wrapped normal distribution is in wide
use, using it here leads to an analysis that is essentially the same as given  in
Section~\ref{sec:GaussRind} for Gaussian noise on $\bbR$. It is appropriate
and interesting 
then to consider the effects of using  the
von Mises distributed in what follows.

Given a set of $N$ points $z_j\in\bbT$, the usual definition of circular mean $\hat z \in\bbT$ is given by
\begin{equation*}
A \hat z = \frac{1}{N}\sum_{j=1}^N z_j
\end{equation*}
The quantity $A$ is a measure of concentration and is related to the circular variance $\rho$ of the sample by $\rho = 1-A^2$.

The von Mises distribution is the  maximal entropy distribution $p(e^{ i \theta})$ subject to the constraint
\begin{equation*}
  \label{eq:5}
  A\mu=\int_{\bbT} e^{i \theta}p(e^{i\theta}) \; d\theta,
\end{equation*}
where $\mu$ is the circular mean and $1-A^2$ is the circular variance. As such, it is the least biased distribution under this constraint. A random variable $Z$ with this distribution is of the form $e^{i\theta_0} Z_0$ where $Z_0$ has circular mean one
and the same circular variance as $Z$. The von Mises density function with circular mean $\mu$ and circular variance $1-A^2$ is
\begin{equation}
\begin{split}
  \label{eq:vonMisesPDF}
  p(z=e^{ i\theta}|\mu, A)&=
\frac{1}{2\pi I_{0}(\kappa)}e^{\kappa \cos(\theta-\mu)}\\
&=\frac{1}{2\pi I_{0}(\kappa)}  e^{\frac{\kappa}{2}(z_{0}^{*}z+z_{0}z^{*})},
\end{split}
\end{equation}
where 
\begin{equation*}
  \label{eq:9}
  A=\frac{I_{1}(\kappa)}{I_{0}(\kappa)},
\end{equation*}
$I_0$, $I_1$ are respectively the first and second modified Bessel functions of the
first kind, and $\bz_0 = e^{i\mu}$.  The value of $A$ determines $\kappa$ via this equation. 

For each edge $e\in E(\Gamma)$, a measurement on $e$ has the form 
\begin{equation*}
  r_e=\omega_e \varepsilon_e
\end{equation*}
where $\bom\in U_*(\Gamma,\bbT)$. Thus, for some 
\begin{equation*}
\bx=(x_v)_{v\in V(\Gamma)}\in C_1(\Gamma,\bbT)  
\end{equation*}
it is possible to write $\omega_e= x_{t(e)} x_{s(e)}^{-1}$. Hence
\begin{equation}
  \label{eq:7}
  r_e=x_{t(e)}\varepsilon_e x_{s(e)}^{-1}
\end{equation}
or, more succinctly,
\begin{equation*}
 \br= \beps\bom =\beps D_* \bx. 
\end{equation*}

\subsection{The estimation problem}
The joint distribution of the noise $\beps\in C_1(\Gamma,\bbT)$ is taken to be 
\begin{equation*}\label{eq:17}
	p(\beps)=\prod_{e\in E(\Gamma)} \frac{1}{2\pi I_0(\kappa_e)}\exp \frac{\kappa_{e}}{2}(\varepsilon_e+\overline{\varepsilon_e}).
\end{equation*}
Consequently, the density of $\br$ conditioned on $\bom$ is
\begin{equation*}
p(\br|\bom)= \prod_{e\in E(\Gamma)}\left(\frac{1}{2\pi I_0(\kappa_e)}\right)\exp \frac{\kappa_{e}}{2}\sum_{e\in E(\Gamma)}  (\overline{\omega_e}r_e+\omega_e \overline{r_e}). 
\end{equation*} 
The log-likelihood is then
\begin{equation}
  \label{eq:19}
  \ell(\br|\bom)=
\sum_{e\in E(\Gamma)}  \frac{\kappa_e}{2}(\overline{\omega_e}r_e+\omega_e\overline{r_e})  +\sum_{e\in E(\Gamma)}\log\frac{1}{2\pi I_0(\kappa_e)}.
\end{equation}

The Fisher information for this estimation problem can be found by first differentiating $\ell(\br|\bom)$ with respect to $\theta_v$ (cf. \eqref{eq:vonMisesPDF}):
\begin{eqnarray*}
  \frac{\partial}{\partial\theta_v}\ell(\br|\bom)
  &=&\sum_{e:s(e)=v} \frac{\kappa_e}{2} i (\overline{\omega_e}r_e-\omega_e\overline{r_e})\\ && \;\;-\sum_{e:t(e)=v} \frac{\kappa_e}{2} i (\overline{\omega_e}r_e-\omega_e\overline{r_e})\\
&=&\Im\Bigl(\sum_{e:t(e)=v} \kappa_e \overline{\omega_e}r_e-\sum_{e:s(e)=v} \kappa_e\overline{\omega_e}r_e\Bigr)\\
&=&\Im \sum_{e\in E(\Gamma)}D_{ve}\kappa_e\overline{\omega_e}r_e
\end{eqnarray*}
Differentiating again with respect to $\theta_u$ yields
\begin{equation*}
 \frac{\partial^2\ell(\br|\bom)}{\partial\theta_u \partial\theta_v} = -\Re\sum_{e\in E(\Gamma)} \kappa_e D_{ue}D_{ve}\overline{\omega_e}r_e
\end{equation*}
The Fisher information is thus
\begin{equation*}
\begin{split}
F_{uv} &= -\sE\left\{ \frac{\partial^2\ell(\br|\bom)}{\partial\theta_u \partial \theta_v}   \right\}\\
  &= \Re \sum_{e\in E(\Gamma)} \kappa_e D_{ue}D_{ve}\overline{\omega_e}\sE\{r_e\}
\end{split}
\end{equation*}
Note that
\begin{equation*}
	\sE\{r_e\} = \frac{I_1(\kappa_e)}{I_0(\kappa_e)}\omega_e,
\end{equation*}
so  that this reduces to
\begin{equation*}
	F = D_W \cF D_W^\sT
\end{equation*}
where $\cF$ is the diagonal matrix 
\begin{equation*}
 \cF = \diag\left(\frac{\kappa_{e_1}I_1(\kappa_{e_1})}{I_0(\kappa_{e_1})}, \cdots,  \frac{\kappa_{e_m}I_1(\kappa_{e_m})}{I_0(\kappa_{e_m})}\right).
\end{equation*}
The quantity $\kappa_e I_1(\kappa_e)/I_0(\kappa_e)$ associated with an edge $e$ may be interpreted as follows. Consider the problem of estimating $\omega_e$ given the datum $r_e$. As above, the conditional density is
\begin{equation*}
	p(r_e | \omega_e) = \frac{1}{2\pi I_0(\kappa_e)}\exp\frac{\kappa_e}{2}(\overline{\omega_e}r_e + \omega_e\overline{r_e}).
\end{equation*}
The parametrization, $\omega_e = \exp(i\phi_e)$, yields the Fisher information
for estimation of $\phi_e$ from $r_e$ as
\begin{equation*}
	\cF_e = -\sE\left\{\frac{\partial^2}{\partial \phi_e^2} \log p(r_e | \omega_e) \right\}=\frac{\kappa_e I_1(\kappa_e)}{I_0(\kappa_e)}
\end{equation*}
Thus when the $\cF_e$ are regarded as edge weights, the graph Fisher information is the weighted Laplacian with these weights. The determinant of the Fisher information is, by Kirchhoff's matrix tree theorem, 
\begin{equation*}
\det F = \sum_S \prod_{e\in S} \cF_e.
\end{equation*}
If the edges of $\Gamma$ share a common $\kappa$, this reduces to
\begin{equation*}
\det F = \frac{\kappa I_1(\kappa)}{I_0(\kappa)} t(\Gamma).
\end{equation*}

\subsection{Maximum-likelihood estimator}
\label{sec:LocalT}

This section considers the problem of determining the maximum-likelihood
estimator $\hat\bx$ for the vertex offsets from data $\br$. Differentiation of
the likelihood \eqref{eq:19} reveals that the critical points must satisfy
\begin{equation*}
\Im \sum_{e\in E(\Gamma)}D_{ve} \kappa_e\overline{\hat\omega_e}r_e = 0
\end{equation*}
at every vertex $v\in V(\Gamma)$. In this expression, $\hat\omega$ denotes the
ML estimate of $\bom$. If the  map $K:C_1(\Gamma,\bbT)\to C_1(\Gamma,\bbC)$ is
defined by 
\begin{equation*}
  K(\beps) = \sum_{e\in E(\Gamma)} \kappa_e \varepsilon_e
\end{equation*}
and the residual is written as $\hat\varepsilon_e = \hat\omega_er_e$, the
critical points of the likelihood satisfy
\begin{align}
  \hat\omega &\in U(\Gamma,\bbT)\nonumber\\
  \Im\left( K(\hat\beps)\right) &\in Z(\Gamma,\bbC)  \label{eq:circkc}
\end{align}
and $\Re\left( K(\hat\beps)\right)$ is the corresponding value of the log-likelihood (up to a constant). These correspond to \eqref{eq:kircest} for the Gaussian case. This condition can be rearranged as
\begin{equation}
  \label{eq:crit}
  \sum_{e\in E(v)}\kappa_{e}\bigl(\overline{\hat\omega_e}r_e \bigr)^{D_{ve}}=\rho_v\in \bbR, 
\end{equation}
for all $v\in V(\Gamma)$. In other words, the weighted sum of the directed
residuals at any vertex must be real. This condition may be seen as a
generalization of the Kirchhoff current law applicable to the group $\bbT$ and
circular statistics. The ML estimate for $\bx$ will be among the solutions of
\eqref{eq:crit}. In contrast to the situation for $\bbR^d$, there are now a
number of critical points. In consequence, to obtain a reliable fast
estimator, it is necessary to distinguish the global maxima from the other
critical points.
\begin{figure}[h]
\begin{center}
\includegraphics[width=\linewidth]{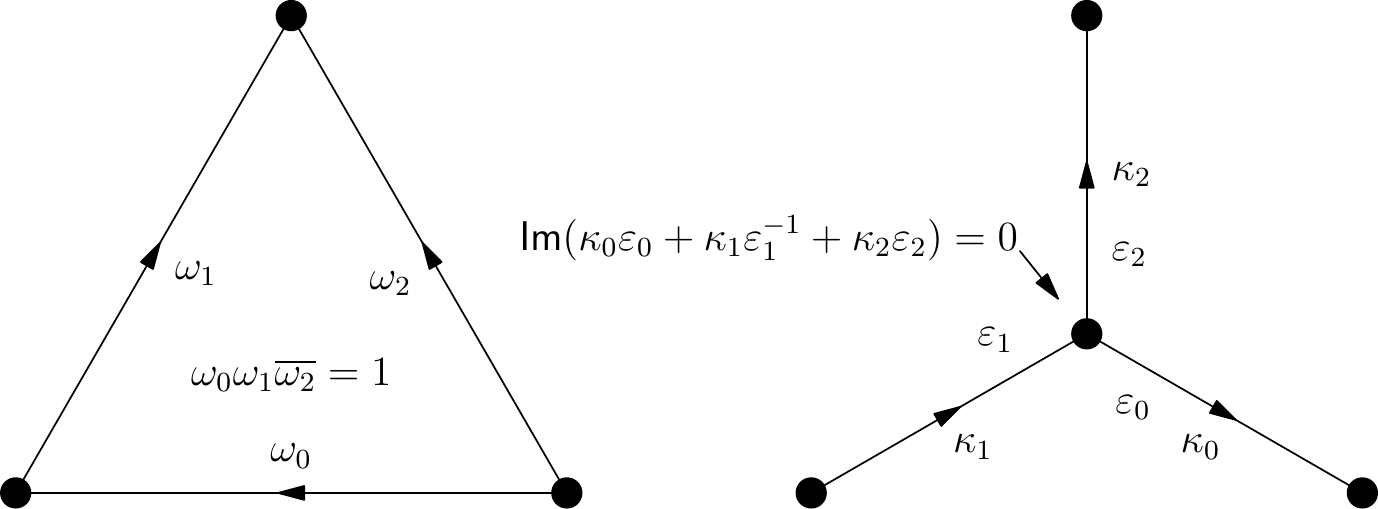}
\end{center}
\caption{Schematic of Kirchhoff laws for the group $\bbT$. At a maximum of the likelihood \eqref{eq:19} the residual $\beps$ satisfies Kirchhoff's current law in the form \eqref{eq:circkc}.}
\end{figure}

Writing 
\begin{equation*}
{\tilde\kappa_v} = \frac{1}{d_v}\sum_{e\in E(v)} \mathcal \kappa_e, 
\end{equation*}
so that $\tilde\kappa_v d_v$ measures the strength of the connection of the
vertex $v$ to the rest of the network, reveals 
that what distinguishes the ML estimate from the other critical points of the likelihood function is that, for moderate noise values,
\begin{equation}\label{eq:mlapprox}
	\frac{1}{d_v}\sum_{e\in E(v)}\frac{\kappa_e}{{\tilde\kappa_v}}\bigl(\overline{\omega_e} r_e\bigr)^{D_{ve}} \approx 1
\end{equation} 

In terms of the augmented adjacency matrix
\begin{equation}
  \label{eq:Qcirc}
  A_{vv'}=
  \begin{cases}
   \kappa_{(v,v')} r_{(v,v')}&\text{ if $(v,v')\in E(\Gamma)$}\\
   \kappa_{(v,v')} r_{(v,v')}^{-1}&\text{ if $(v',v)\in E(\Gamma)$}\\
    0&\text{ otherwise}
  \end{cases},
\end{equation}
equation \eqref{eq:crit} can be reformulated with the aid of \eqref{eq:7} as
\begin{equation}\label{eq:Qgeneig}
  Q\bx = N^{-1}A\bx = G\bx,
\end{equation}
where $N_{v,v'} = \tilde\kappa_v d_v\delta_{v,v'}$ and $G$ is a real diagonal matrix. The approximation \eqref{eq:mlapprox} corresponds to $\tilde\kappa_v d_v \approx \rho_v$ or $G\approx\bbI$. When $\kappa_e =\kappa$ for all $e\in E(\Gamma)$, the matrix $N$ is $\kappa$ times the degree matrix. 

A fast estimator for $\bx$ may be obtained as follows. First, find the eigenvector $\by\in\bbC^n$ corresponding to the largest eigenvalue of $Q$, 
\begin{equation}\label{eq:fastest}
	Q\by=\lambda\by
\end{equation}
Then $\hat x_v = y_v/\abs{y_v}$ for all $v\in V(\Gamma)$. 
The structure of this  estimator is elucidated  by consideration of  its local implementation,
which can be obtained by applying the power method \cite{Horn1993} to
\eqref{eq:fastest}. This involves  application of  the matrix $Q$ repeatedly to a
random initial vector $\by^{(0)}$. The method results in convergence  to an eigenvector
of $Q$ corresponding to the eigenvalue of largest magnitude provided this is
the only eigenvalue of largest magnitude and the initial $\by^{(0)}$ is not
orthogonal to a left eigenvector of $Q$. In practice, to avoid numerical
overflow, the resultant vector is re-normalized after each application of
$Q$. This is a critical point in turning \eqref{eq:fastest}, or any other
eigenvector based estimator, into an algorithm that can be run locally on the
network. The normalization of $\by$ is a global operation since it involves
each vertex knowing the values of $\by$ across the entire network. Without the
ability to normalize vector, it is necessary to ensure that the eigenvalue of
largest magnitude is close enough to unity that the iteration converges well
before an overflow or underflow condition is reached. The algorithm proposed
here indeed has the property that the largest magnitude eigenvalue $\lambda$
is approximately one.  Further, as demonstrated in the simulations below, it
operates reliably to align the network without renormalization.

To examine the local operation of this algorithm more closely, write $y_v=a_v x_v$ where $a_v$ is a real amplitude and $x_v\in\bbT$ for each $v\in V(\Gamma)$. In addition to its phase $x_v\in\bbT$, each vertex keeps a circular variance $0\leq a_v \leq 1$ which measures the degree to which it agrees with its neighbors. The local update rule for the case $\kappa_e =\kappa$ for all $e\in E(\Gamma)$ is 
\begin{equation}\label{eq:localQeig}
  a_v^{(m+1)}x_v^{(m+1)} = \frac{1}{d_v}\sum_{u\sim v} a_u^{(m)}\;(x_u^{(m)}r_{(u,v)})^{D_{v,(u,v)}}
\end{equation}
Thus, at each update, the vertex $v$ resets to the weighted circular mean of what its nearest neighbors predict its phase should be (compare this to Jacobi's method for alignment in $\bbR^d$). The weighting is based on how well each of the neighbors are aligned with their own neighbors. In this way, the contributions of vertices that have not yet converged are discounted relative to the ones from vertices that have converged to alignment with their neighbors. More generally, the update takes the form
\begin{equation*}
  a_v^{(m+1)}x_v^{(m+1)} = \frac{1}{d_v}\sum_{u\sim v} \frac{\kappa_{(u,v)}}{\tilde\kappa_v} a_u^{(m)}\;(x_u^{(m)}r_{(u,v)})^{D_{v,(u,v)}}
\end{equation*}

The convergence rate of the power method algorithm depends on the distance between the largest and next-to-largest magnitude eigenvalues. The Gershgorin circle theorem \cite{Horn1993} implies that
\begin{equation*}
  -1 \leq \lambda \leq 1
\end{equation*}
The impediment to convergence comes from eigenvalues of $Q$ near $-1$. Regularization can alleviate this to some extent. Consider the matrix
\begin{equation*}
Q_\beta = (N+\beta I)^{-1} (A+\beta \bbI) 
\end{equation*}
where $\beta\geq 0$ is a regularization parameter. If $\by$ satisfies
\begin{equation*}
  Q_\beta\by = \lambda\by
\end{equation*}
then
\begin{equation*}
  Q\by = \lambda\by + \beta (\lambda-1)N^{-1}\by \approx \lambda\by
\end{equation*}
Thus, if the maximum magnitude eigenvalue is close to one, the regularization leaves it close to 1. For the regularized eigenvalue problem, Gershgorin's theorem implies
\begin{equation*}
  -1 + \frac{2\beta}{\tilde\kappa_{\max} + \beta} \leq \lambda \leq 1
\end{equation*}
where $\tilde\kappa_{\max}$ is the largest element of the matrix $N$. Thus the parameter $\beta$ can be used to bound the smallest eigenvalue away from $-1$, the effectiveness being mitigated by any vertex having a large value of $\tilde\kappa_v$.
  

This estimator exhibits remarkably good performance. In cases tested, the estimator gives a mean circular error performance indistinguishable from that of actual maximum likelihood, even when $\kappa\approx 1$. Although it seems to have no real bearing on practical estimation performance, it is possible to construct estimators that give values closer to the maximum-likelihood estimate. One example of such an estimator, the hybrid maximum-likelihood algorithm, starts by running a suitable number of iterations of the local algorithm just described. Then it switches to the following iteration 
\begin{equation}\label{eq:localML}
  a_v^{(m+1)} x_v^{(m+1)} = \frac{1}{d_v}\left(\sum_{u\sim v} \frac{\kappa_{(u,v)}}{\tilde\kappa_v^{(m)}} a_u^{(m)}(x_u^{(m)}r_e)^{D_{ve}}\right),
\end{equation}
where
\begin{equation*}
  \tilde\kappa_v^{(m)} =  \frac{1}{d_v}\Re \left( (A \bx^{(m)})_v/x^{(m)}_v \right). 
\end{equation*}
At a particular vertex, the algorithm stops when the size of the imaginary component of $(A \bx^{(m)})_v/x^{(m)}_v$ drops below some threshold. At this point, $\bx$ satisfies the condition \eqref{eq:Qgeneig} with tolerance corresponding to the threshold, and thus a maximum of the likelihood function to this tolerance is obtained.

\begin{figure}[h]
  \begin{center}
    \includegraphics[width=0.7\linewidth]{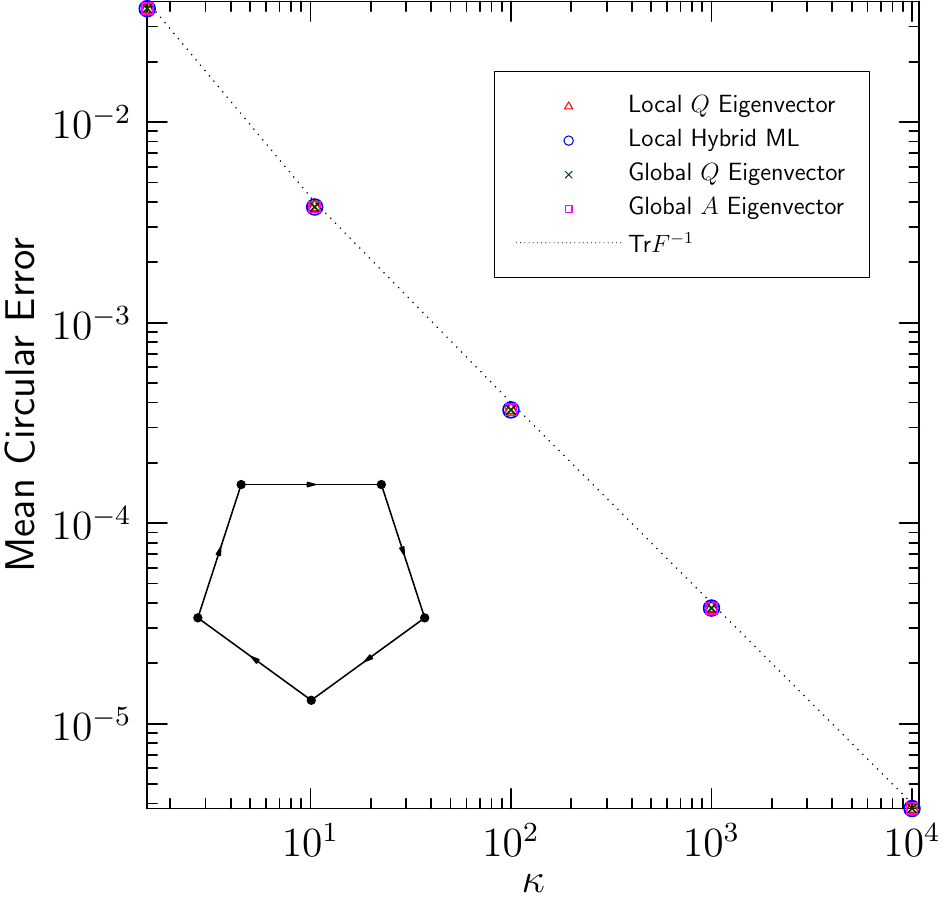}
    \caption{Mean circular error versus concentration parameter $\kappa$ for the four estimators operating on the five-node network shown in the inset.}
    \label{fig:sim1}
  \end{center}
\end{figure}

\begin{figure}
  \begin{center}
    \includegraphics[width=0.7\linewidth]{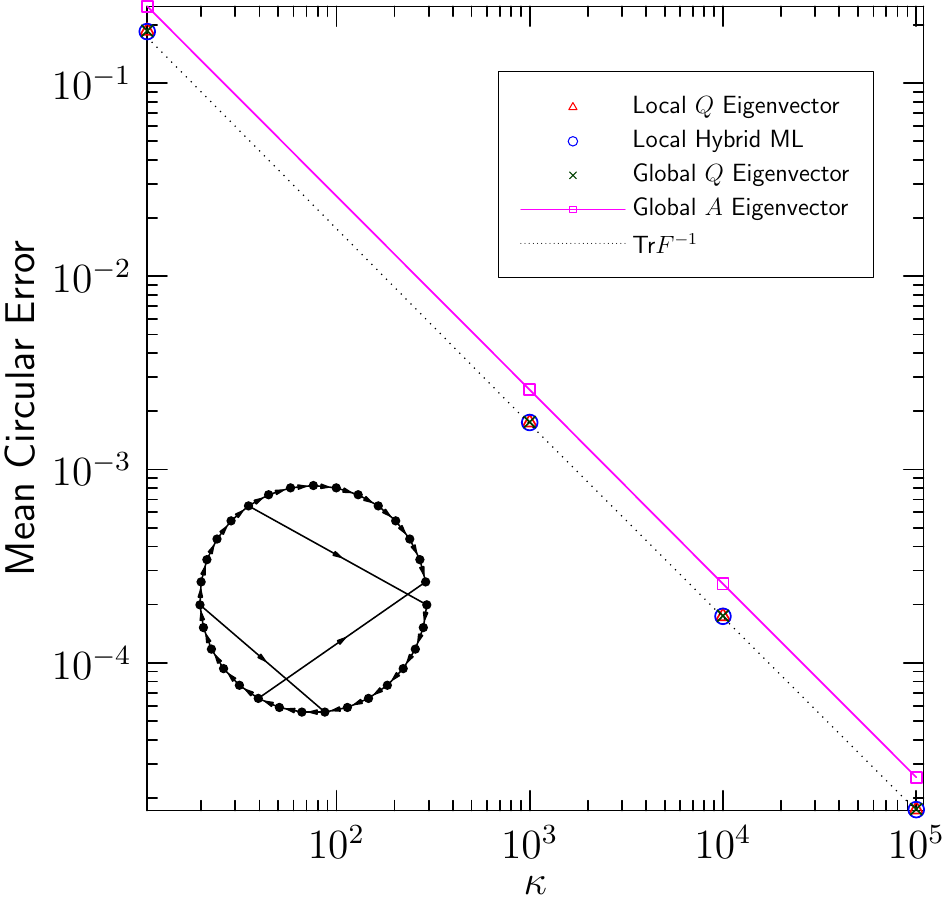}
    \caption{Mean circular error versus concentration parameter $\kappa$ for the four estimators operating on the $31$-node network shown in the inset.}\label{fig:sim2}
  \end{center}
\end{figure}

The algorithms described above were tested on two simulated networks, one with five nodes and the other with $31$ nodes. The errors on each of the edges were identically distributed with concentration parameter $\kappa$. Two global algorithms were tested the where the eigenvectors corresponding to the largest eigenvalues of the matrices $Q$ and $A$ defined in \eqref{eq:Qcirc} and \eqref{eq:Qgeneig} were computed. These were compared with the two local algorithms; i.e., the algorithm \eqref{eq:localQeig} for arriving at an the eigenvector of $Q$ corresponding to the largest eigenvalue and the hybrid maximum-likelihood algorithm described in \eqref{eq:localML}. These local algorithms were implemented on a network simulation with only nearest neighbor communication and with purely local stopping criteria. The results are shown in Figures~\ref{fig:sim1} and \ref{fig:sim2}, where the network structure is shown in the bottom left hand corner of each graph. The figures show the mean circular error across the network; i.e., 
\begin{equation*}
  CE(\hat \bx) = 1 - \Bigl|\frac{1}{|V(\Gamma)-1|}\sum_{v\in V(\Gamma)/v_0}\hat x_v \overline{x}_v\Bigr|^2
\end{equation*}
where $x_v$ is the true value of the node offset. In each case, these results are compared with the the trace of the inverse of the Fisher information, 
\begin{equation*}
  \tr F^{-1} = \frac{I_0(\kappa)}{\kappa I_1(\kappa)} \tr{\tilde L}.
\end{equation*}
These results indicate that the local $Q$ eigenvector estimator works as well as the global methods, gives essentially maximum-likelihood performance, and lies very close to the trace of the inverse Fisher information.

\section{Abelian Lie Groups}
\label{sec:abelian_Lie}

This section briefly describes the situation in which the measurements reside in a general connected abelian Lie group. Such groups are just products of groups considered in the previous sections; i.e., $G$ is of the form $\bbR^d\times\bbT^q$, and its elements may thus be written as $\bg=(\bx,\bz)$ where $\bx\in\bbR^d$ and $\bz=(e^{2\pi i\theta_{k}})_{k=1}^q$. Its dual group, which is the group of homomorphisms from $G$ to $\bbT$, is identifiable with $\bbR^d\times\bbZ^q$, with elements $\btau=(\by,\bn)$, via the pairing
\begin{equation*}
  \label{eq:pairing_general}
\langle\bg,\btau\rangle=\langle (\bx, \bz),(\by,\bn)\rangle =\exp i\bigl(\bx.\by+
2\pi\sum_{k=1}^{q}n_{k}\theta_{k}\bigr). 
\end{equation*}
As in the formulations treated previously, the edges of the graph $\Gamma$ are labeled with offsets that are elements of $G$. Rather than treating this case in complete detail, the focus of what follows in on synopsis of the aspects of the problem where significant modification of the approach described for earlier cases is needed to treat this more general situation.

A key issue in this setting is the choice of a suitable model for the
distribution of the noise corrupting the edge labels. Little work has
been done on probability distributions on abelian Lie groups. Even the
situation of a $q$-torus, which is simply a product of circles, for $q>2$ is
little explored, although the case $q=2$ is addressed in work of Singh
\emph{et al.}\cite{Singh2002}, in which a version of the von~Mises
distribution is used in connection with an application to estimation of
torsion angles in complex molecules. Specifically, they use
\begin{equation}
\label{eq:singhetal}
\begin{split}
p(\theta_1,\theta_2) & = C\exp\bigl(\kappa_1\cos(\theta_1 - \mu_1) \\
& \quad + \kappa_2\cos(\theta_2 - \mu_2) \\
& \quad + \lambda\sin(\theta_1-\mu_1)\sin(\theta_2-\mu_2)\bigr).
\end{split}
\end{equation}
The authors have studied this topic and intend to address it in a later paper, where a detailed analysis of appropriate error distributions on such groups will be presented.  For the present purpose, however, it suffices to assume sufficiently smooth densities $p_e(g)=p_e(\bx,\bz)$ for $g=(\bx,\bz)\in G=\bbR^d \times\bbT^q$ and $e\in E(\Gamma)$ to allow the Fisher information to be calculated. Also, for this discussion, attention is restricted to the case where noise on distinct edges is independent. 

The methods adopted in the previous examples can be mimicked and generalized
as follows, where now the group operation is written additively 
(so that the circle group $\bbT$ is represented as the real numbers modulo
$2\pi$).  The data vector $\br$ is, therefore,  given by
\begin{equation*}
  \begin{split}
  \br_e &= \bom_e + \beps_e\\
      &= \bx_{t(e)}-\bx_{s(e)} + \beps_e
  \end{split}
\end{equation*}
for each $e\in E(\Gamma)$. In this expression, $\bom$ is parameterized in terms of the vertex offsets $\bx$. The probability density for $\beps$ is
\begin{equation*}
  f(\beps) = \prod_{e\in E(\Gamma)} f_e(\beps_e). 
\end{equation*}
The density for the noisy measurements is 
\begin{equation*}
  f(\br|\bx) = \prod_{e\in E(\Gamma)} f_e(\br_e-\bx_{t(e)}+\bx_{s(e)}) 
\end{equation*} 
and, for given data $\br$, the log-likelihood function is 
\begin{equation*}
  \label{eq:27}
 \ell(\br|\bx)  = \sum_{e\in E(G)} \log f_e(\br_e-\bx_{t(e)}+\bx_{s(e)})   
\end{equation*}
where $\br_e$ and $\bx_e$ are elements of $G$. It remains to define an ``edge Fisher information'' $\cF_e$ as
\begin{equation*}
  \label{eq:edge_fisher}
\cF_e=-\sE\bigl\{\bnabla_{\bom_e}^2\log p_e(\br_e-\bom_e)\bigr\}
\end{equation*}
where $\bnabla^2$ involves partial derivatives in each of the real and circular coordinates of $G$. Observe that $\cF_e$ is a $(d+q)\times(d+q)$ matrix indexed by the coordinates of $G$. 

By analogy to the case of $\bbR^d$, the Fisher information matrix for $\bx$ is given by
\begin{equation*} 
\begin{split}
 F^W &= -\sE\left\{\bnabla^2_{\bx} \log p(\br | \bom) \right\} \\
&= \left(\bnabla_{\bx}\bom\right)^\sT\cF\bnabla_{\bx}\bom
\end{split} 
\end{equation*}
where $\cF$ is a $m(d+q)\times m(d+q)$ matrix ($m$ is the number of edges in the graph) with components
\begin{equation*}
  \cF_{(e,s),(e',t)} = \delta_{e,e'} [\cF_e]_{s,t}
\end{equation*}
for $s,t = 1,\cdots, d+q$. Using 
\begin{equation*}
\bnabla_{\bx}\bom = \bbI\otimes D_W^\sT,
\end{equation*}
gives
\begin{equation*} 
 F^W = (\bbI\otimes D_W)\cF(\bbI\otimes D_W^\sT).
\end{equation*}
In a way that closely parallels the multi-dimensional Gaussian case treated in Section~\ref{sec:Rdgauss} (cf. equation \eqref{eq:Rdfisherind}), the determinant of the Fisher information in this case is
\begin{equation*} 
 \det F^W = \sum_{\cS} \det(\cF_{\cS\cS}) ,
\end{equation*}
where the sum is over all multi-spanning trees $\cS=(S_1,S_2,\cdots,S_{d+q})$ consisting of one spanning tree for each dimension of $G$.

\section{Discussion and Conclusions}
\label{sec:wrapup}

The preceding sections have introduced a statistical framework for registration and synchronization of data collected at the nodes of a network.  The approach, which formulates the registration problem as one of optimally estimating true offsets between data at communicating nodes from noisy measurements of these values. Explicit estimators were derived and analyzed for the case in which the data reside in $\bbR^d$ and the noise corrupting the measurements is Gaussian. These solutions were pointed out to have a homological character leading to conditions akin to Kirchhoff's laws that optimal estimates and their residual errors must satisfy. They were shown to provide insight about how network topology can be designed and adapted to promote accurate synchronization across the network.  

The phase alignment problem in which the data are on the circle $\bbT$ has also be treated explicitly and has been pointed out to manifest the critical properties of the more general case in which the data belong to a connected abelian Lie group.  Further, iterative local algorithms, in which nodes only make use of information from their nearest neighbors in each iteration, have been described and empirically demonstrated.

Important practical cases, including alignment of local coordinate systems, involve non-abelian Lie groups.  Application of the approach set forth here to this class of problems will be treated in a sequel to this paper.

\nocite{*}
\bibliographystyle{IEEEtran}
\bibliography{graph_est_refs}

\end{document}